\documentclass[journal]{IEEEtran}
\ifCLASSINFOpdf
   \usepackage[pdftex]{graphicx}
\else
   \usepackage[dvips]{graphicx}
\fi
\usepackage{caption}
\usepackage{amsmath}
\usepackage{hyperref}
\usepackage{array}
\usepackage[flushleft]{threeparttable}

\usepackage{tabu}
\usepackage{multirow}

\usepackage{algorithm}  
\usepackage{algpseudocode}  

\DeclareMathOperator*{\argmin}{arg\,min}  
\begin{document}
%
\title{Bayesian Convolutional Neural Networks for Compressed Sensing Restoration}

%
%
%

\author{Xinjie~Lan,~\IEEEmembership{Student Member,~IEEE,}
        Xin~Guo,~\IEEEmembership{Student Member,~IEEE,}
        Kenneth~E.~Barner,~\IEEEmembership{Fellow,~IEEE,} 
        
\thanks{This material is based upon work supported by the National Science Foundation under Grant No. 1319598. The material in this paper was presented in part at IEEE GlobalSIP, Montreal, Canada, November 2017.}
\thanks{Xinjie Lan, Xin Guo, and Kenneth E. Barner are with Department of Electrical and Computer Engineering, University of Delaware, Newark, DE, 19711 USA (email: lxjbit, guoxin, barner@udel.edu)}
}

\maketitle

\begin{abstract}
Deep Neural Networks (DNNs) have aroused great attention in Compressed Sensing (CS) restoration.
However, the working mechanism of DNNs is not explainable, thereby it is unclear that how to design an optimal DNNs for CS restoration.
In this paper, we propose a novel statistical framework to explain DNNs, which proves that the hidden layers of DNNs are equivalent to Gibbs distributions and interprets DNNs as a Bayesian hierarchical model.
The framework provides a Bayesian perspective to explain the working mechanism of DNNs, namely some hidden layers learn a prior distribution and other layers learn a likelihood distribution.
Moreover, the framework provides insights into DNNs and reveals two inherent limitations of DNNs for CS restoration.
In contrast to most previous works designing an end-to-end DNNs for CS restoration, we propose a novel DNNs to model a prior distribution only, which can circumvent the limitations of DNNs.
Given the prior distribution generated from the DNNs, we design a Bayesian inference algorithm to realize CS restoration in the framework of Bayesian Compressed Sensing.
Finally, extensive simulations validate the proposed theory of DNNs and demonstrate that the proposed algorithm outperforms the state-of-the-art CS restoration methods.
\end{abstract}


%
\IEEEpeerreviewmaketitle

\section{Introduction}
%
%
%
%
\IEEEPARstart{S}{ampling} theory is a cornerstone of signal processing, which indicates that a band-limited signal can be precisely recovered from its uniform samples as long as the sampling rate is at least twice the bandwidth of this signal, namely the Nyquist rate \cite{Shannon}. 
However, the exponential growth of the Nyquist rate makes signal processing extremely difficult and complicated in the big data era. 
In other words, conventional sampling theory plagues our ability to acquire, transmit, and process high dimensional dataset, thereby spurring tremendous interest in novel sampling theories and techniques.

Compressive Sensing (CS) provides an alternative paradigm for sampling theory \cite{Candes, Donoho}.
In general, a $K$-sparse signal $\boldsymbol{x} \in \boldsymbol{R}^{N}$ is to be sampled by a linear measurement matrix $\boldsymbol{A}  \in R^{M \times N} (M < N)$, and the measurement $\boldsymbol{y}$ can be written in matrix form as $\boldsymbol{y = Ax + n}$, where $\boldsymbol{n} \in R^{M}$ is commonly assumed as Gaussian noise with variance $\sigma_n^2$. 
If $\boldsymbol{A}$ satisfies \textit{Restricted Isometry Property} (RIP), CS guarantees that $\boldsymbol{x}$ can be accurately restored from $M = \mathcal{O}(K log(N/K)$ measurements with high probability \cite{Blumensath, RIP}. 
\pagebreak


Among all traditional methods of CS restoration, Bayesian Compressed Sensing (BCS) has received much attention, since it can easily take into account the prior knowledge of $\boldsymbol{x}$, namely the inherent signal structures of $\boldsymbol{x}$, to improve CS restoration by specifying a prior distribution \cite{Duarte, CoSaMP-tree, BCS}.
In the context of BCS,  CS restoration is reformulated as a Bayesian posterior inference problem, in which the signal structure is expressed by a prior distribution $p(\boldsymbol{x}; \boldsymbol{\theta})$ and the likelihood distribution of CS is commonly formulated as a Gaussian distribution \cite{laplace_bcs, bp_bcs}.
After deriving the posterior distribution of CS given the prior and likelihood distributions, we can realize CS restoration by inferring the posterior distribution.

Since $p(\boldsymbol{x}; \boldsymbol{\theta})$ plays an important role in describing the signal structure of interest, the key of BCS is to choose a $p(\boldsymbol{x}; \boldsymbol{\theta})$ for modeling the signal structure precisely and comprehensively. 
Lots of prior distributions have been used to model various signal structures in the framework of BCS, e.g., Laplace distribution \cite{laplace_bcs}, Markov Random Fields (MRFs) \cite{CS-markov-tree, Cevher, Schniter}, Spike and Slab model \cite{BCS-tree,BCS-neig}, Gaussian Mixture Model (GMM) \cite{CS-GSM, GM-AMP, Yanting, xinjie}. 
However, the above prior distributions have two limitations. 
First, they commonly exploit hand-crafted linear filters, which are difficult to express the signal structure precisely unless taking lots of trials to modify the parameters of the filters. 
Second, they merely model the signal structure in a small neighborhood in order to reduce the complexity of parameter optimization, but most signal structures have abundant high-order dependencies \cite{GSM-denoise}. 
As a result, they are difficult to describe the signal structure comprehensively as well.

As the resurgence of Deep Neural Networks (DNNs) in various applications like speech recognition \cite{ANN-GMM} and image classification \cite{CNN}, 
DNNs have aroused considerable attention in CS restoration. 
Compared to traditional BCS algorithms, DNNs can precisely describe multiple signal structures based on their powerful representation ability \cite{representation-dl} and efficient training algorithms \cite{backpropagation}. 
More specfically, some works attempt to design an end-to-end DNNs to directly restore $\boldsymbol{x}$ from $\boldsymbol{y}$ for CS restoration \cite{ReconNet, DR2, MRI-cnn}. 
Meanwhile, other works combine traditional CS restoration and DNNs to realize CS restoration \cite{Learninginvert, LDAMP}.
It is noteworthy that some DNNs present better performance than traditional BCS methods \cite{DR2, LDAMP}. 

However, a fundamental problem of DNNs is that they are vulnerable to perturbation \cite{CNNs_fooled, CNNs_fooled1, CNNs_robust}, which results in bad CS restoration in noisy situation.
Moreover, all existing theories cannot convincingly clarify the internal logic of deep learning and DNNs have been viewed as "black boxes" \cite{DNN_blackbox}.
That means we still don't know how to design an optimal neural networks for CS restoration.
\pagebreak

In this paper, we attempt to shed light on the above problems and propose a novel algorithm for CS restoration based on DNNs.
Overall, our contributions to this literature are three folds. 
First, we propose a novel statistical framework to explain the architecture of DNNs, which proves that the hidden layers of DNNs are equivalent to Gibbs distributions and interprets DNNs as a Bayesian hierarchical model.
In particular, the proposed framework provides a novel Bayesian perspective to explain DNNs, namely some hidden layers are used to learn a prior distribution $p(\boldsymbol{X})$ and the remaining layers are responsible for learning a likelihood distribution $p(\boldsymbol{\hat{Y}|X})$, where $\boldsymbol{X}$ and $\boldsymbol{\hat{Y}}$ denote the random variables of the input and output of DNNs, respectively.

Second, we provide insights into DNNs in four aspects: 
(i) the application scope of DNNs is confine to the situation that the training dataset provides great or equal to the information of the training labels;
(ii) we cannot achieve the state-of-the-art performance by simply applying DNNs to CS restoration.
(iii) the activation function of DNNs cannot denoise very well, which is an important reason for the vulnerability of DNNs. 
(iv) we unify traditional BCS methods and DNNs and provide an overall picture of all BCS algorithms;

Third, we propose a novel CS restoration algorithm based on DNNs.
In contrast to design an end-to-end DNNs for CS restoration, we propose a novel Convolutional Neural Networks (CNNs) to model a prior distribution $p(\boldsymbol{x}; \boldsymbol{\theta})$ only. 
In particular, the proposed CNNs have a dual property: 
(i) it has a closed-form expression as traditional prior distributions;
(ii) it preserves the same powerful representation ability as DNNs.
The duality enable the proposed CNNs to overcome the limitations of BCS and DNNs simultaneously.
Given the prior distribution $p(\boldsymbol{x}; \boldsymbol{\theta})$ corresponding to the CNNs and the likelihood distribution of CS, we derive the posterior distribution of CS and propose a novel Bayesian inference algorithm to realize CS restoration. 

This paper is organized as follows. 
Section II lays out the necessary background related to this topic. 
We describe the proposed statistical framework to explain DNNs in Section III. 
Subsequently, Section IV proposes the insights into DNNs.
Section V presents the proposed CS restoration algorithm.
In the end, numerical simulations validate our proposed theory and algorithm in Section VI.

\begin{figure*}[t]
\centering
\centering\includegraphics[scale=0.26]{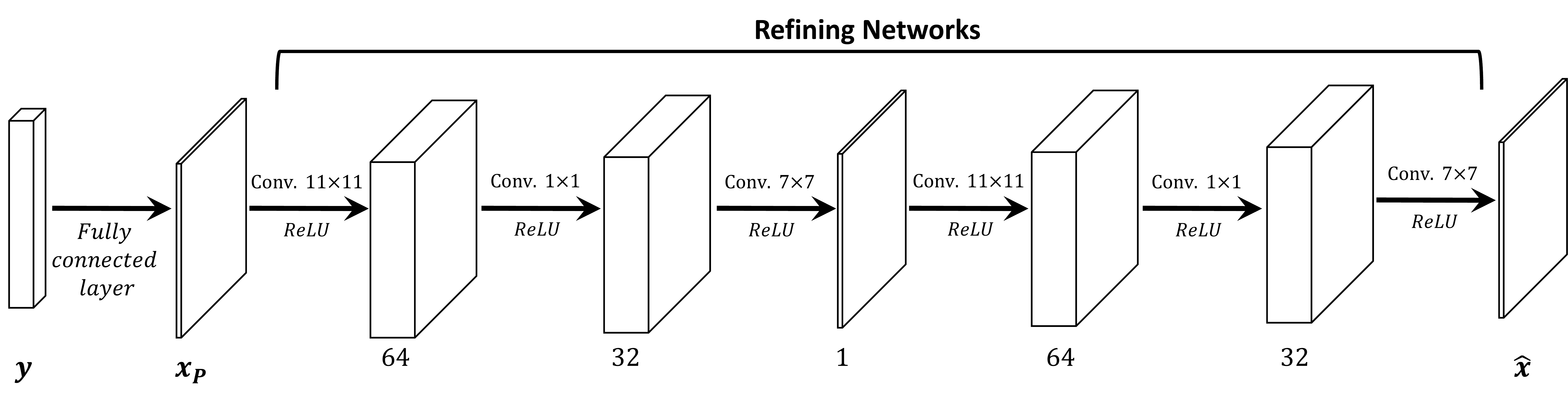}
\caption{The architecture of DNNs for CS restoration.
This figure also illustrate an existing DNNs for CS restoration, namely ReconNet \cite{ReconNet}, which uses a single fully connected layer to obtain $\boldsymbol{x}_P$ and six convolutional layers (abbr. Conv.) to refine $\boldsymbol{x}_P$.
The third block represents a convolutional layer of ReconNet consist of 64 linear filters with dimension $11 \times 11$.
}
\label{inverse_cnns_layers}
\end{figure*}

\section{Background}

This section reviews three fundamental topics related to our research: Gibbs distribution, BCS, and DNNs.

\subsection{Gibbs Distribution}
\label{gibbs_section}

In statistical mechanics, a Gibbs distribution is a probability measure that gives the probability of a certain state as a function of the state's energy and system temperature, which can be expressed as
\begin{equation} 
\label{Gibbs} 
{\textstyle
p_k(\boldsymbol{x}; \boldsymbol{\theta}) = \frac {1}{Z(\boldsymbol{\theta})}\text{exp}[-\frac{1}{T}g_k(\boldsymbol{x; \theta_k})]
}
\end{equation}
where $\boldsymbol{x}$ indicate all samples in a system, $g_k(\boldsymbol{x; \theta_k})$ is the \textit{energy function} of state $k$, and $T$ is system temperature \cite{Geman}.
In addition, $Z(\boldsymbol{\theta}) = \sum_{k=1}^K {\text{exp}(-g_k(\boldsymbol{x; \theta})/T)}$ is the partition function, $K$ denotes the number of all states in the system of interest, and $\boldsymbol{\theta} = \bigcup_k{\boldsymbol{\theta}_k}$ indicate all parameters.

It is important to note that Gibbs distribution is equivalent to Markov Random Fields (MRFs), which can be formulated below \cite{MRF_IA, GibbsPiror}. 
\begin{equation} 
\label{Gibbs_MRF} 
{\textstyle
p(\boldsymbol{x}; \boldsymbol{\theta}) = \frac {1}{Z(\boldsymbol{\theta})}\text{exp}[-\frac{1}{T}g(\boldsymbol{x}; \boldsymbol{\theta})]
}
\end{equation}
In terms of MRFs, the \textit{energy function} $g(\boldsymbol{x}; \boldsymbol{\theta})$ is defined as
\begin{equation} 
\label{energy} 
{\textstyle
g(\boldsymbol{x}; \boldsymbol{\theta}) = \sum_kf_k^{NL}(f_k(\boldsymbol{x}; \boldsymbol{\theta}_k))
}
\end{equation}
where $f_{k}^{NL}(f_k(\cdot))$ represents a clique potential function, and $f_{k}^{NL}$ is an non-linear function, $f_k(\cdot)$ is a linear filter to describe local dependence \cite{GibbsPiror}. 
The partition function $Z(\boldsymbol{\theta})$ is defined as $\sum_{\boldsymbol{x}}{\text{exp}(-g(\boldsymbol{x; \theta})/T)}$ in the context of MRFs.

Gibbs distribution, especially MRFs, is a commonly used prior distribution in BCS field, since it can  easily derive an arbitrary distribution to model the signal structure of interest by redefining $g(\boldsymbol{x}; \boldsymbol{\theta})$ \cite{CS-markov-tree, Cevher, Schniter, BCS-tree}.
Meanwhile, Gibbs distribution has various connections with DNNs \cite{MRFs_MLP, MRF_RNN}.
First, it has been proven that softmax layer is identical to Gibbs distribution \cite{Gibbs_softmax, energy_learning}.
Second, some works demonstrate that a convolutional layer with an non-linear layer can be formulated as MRFs model \cite{CRF_RNN, Xinjie2}.

\subsection{Bayesian Compressed Sensing}
\label{lim_bcs}

Given the expression of CS, namely $\boldsymbol{y = Ax + n}$, BCS converts CS restoration into a Bayesian posterior inference problem \cite{BCS, laplace_bcs, bp_bcs}. 
Since the noise $\boldsymbol{n}$ is commonly assumed as a Gaussian distribution with unknown variance $\sigma_n^2$, the likelihood distribution of CS can be formulated below. 
\begin{equation} 
\label{likelihood} 
{\textstyle
p(\boldsymbol{y}|\boldsymbol{x}, \boldsymbol{A}, \sigma_n^2) = \frac{1}{\sqrt{(2\pi\sigma_n^2)^{N}}} \text{exp}(-\frac{1}{2\sigma_n^2} \cdot \Vert \boldsymbol{Ax} - \boldsymbol{y} \Vert^2_2)
}
\end{equation}

Assuming the prior distribution is expressed as a Gibbs distribution $p(\boldsymbol{x}; \boldsymbol{\theta})$ to describe the signal structure of interest,
the posterior distribution $p(\boldsymbol{x}|\boldsymbol{A},\boldsymbol{y}, \sigma_n^2)$ for CS restoration can be derived as follows given the above prior and likelihood distributions based on Bayes' theorem.
\begin{equation} 
\label{posterior} 
p(\boldsymbol{x}|\boldsymbol{A},\boldsymbol{y}, \sigma_n^2) \propto p(\boldsymbol{y}|\boldsymbol{x}, \boldsymbol{A}, \sigma_n^2) \cdot p(\boldsymbol{x}; \boldsymbol{\theta}) 
\end{equation}

Moreover, the solution to CS restoration can be derived below based on the Maximum a Posteriori (MAP) principle.
\begin{equation} 
\label{MAP} 
\boldsymbol{\hat{x}}_{\text{BCS}} = \argmin_{\boldsymbol{x} \in R^N} \Vert \boldsymbol{Ax} - \boldsymbol{y} \Vert^2_2 + \frac{2\sigma_n^2}{T} \cdot g(\boldsymbol{x}; \boldsymbol{\theta})
\end{equation}

The above equation suggests that $\boldsymbol{\hat{x}}_{\text{BCS}}$ crucially depends on $g(\boldsymbol{x}; \boldsymbol{\theta})$, which describes the signal structure of interest.
However, an unsolved problem of $g(\boldsymbol{x}; \boldsymbol{\theta})$ is to derive optimal linear filters $f_k(\cdot)$ for modeling the signal structure precisely and comprehensively.
Initially, hand-crafted linear filter $f_k(\cdot)$ is directly utilized to model signal structure \cite{BCS-tree, BCS-neig}, which needs lots of trials to obtain good CS  restoration.
Expectation-Maximization (EM) method is commonly explored to optimize $f_k(\cdot)$, but it still cannot model the signal structure precisely since it requires too many hyper-parameters \cite{GM-AMP}.
Meanwhile, existing methods only describe signal structure in a small neighborhood, but most signal structures have abundant high-order dependencies \cite{GSM-denoise}, which means that $f_k(\cdot)$ is hard to simulate the signal structure comprehensively.

\subsection{Deep Neural Networks for CS restoration}
\label{dnns_section}

\begin{figure}[b]
\centering
\includegraphics[scale=0.2]{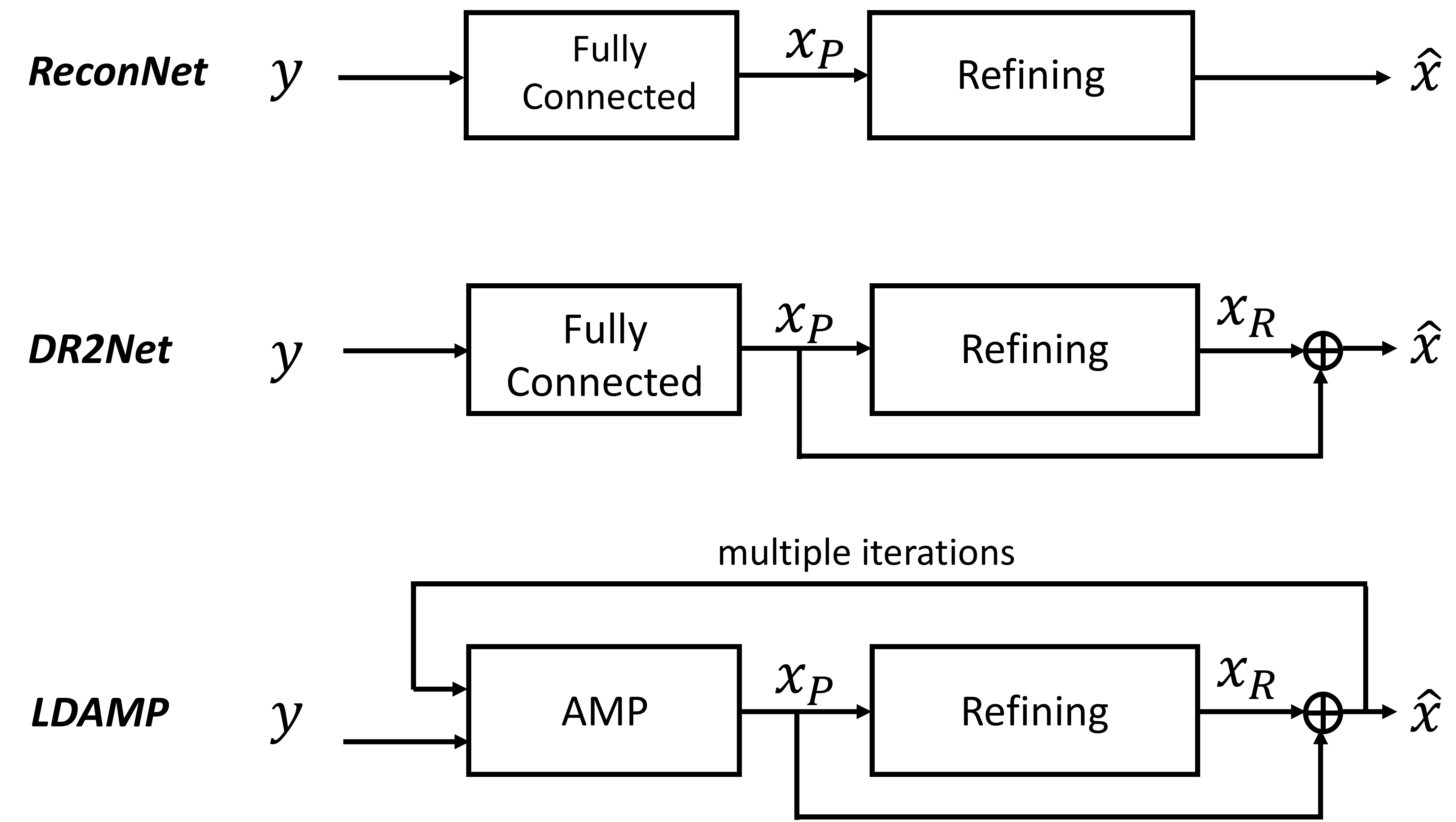}
\caption{\small{The architectures of three neural networks for CS restoration, $\boldsymbol{y}$ denotes a CS measurement, and $\hat{\boldsymbol{x}}$} is the restored signal.}	
\label{threarchitectures}
\end{figure}

Because of their powerful representation ability \cite{representation-dl} and efficient training algorithms \cite{backpropagation}, DNNs have the potential to solve the aforementioned problem of BCS. 
Consequently, numerous efforts recently have been devoted to applying DNNs to CS restoration field \cite{Learninginvert, ReconNet, DR2, MRI-cnn, LDAMP}. 

Since standard DNNs can be viewed as a function projecting the high dimensional input, such as an image dataset, to the low dimensional output like classification labels, 
the most intuitive way of designing DNNs for CS restoration is to reverse standard DNNs as Fig. \ref{inverse_cnns_layers}.
More specifically, the DNNs first use fully connected layers to derive a preliminary restoration $\boldsymbol{x}_P$, and take advantage of the powerful representation ability of convolutional layers to refine $\boldsymbol{x}_P$ for deriving the final restoration $\hat{\boldsymbol{x}}$. 
In particular, the sub-networks for refining $\boldsymbol{x}_P$ is called refining networks here.

Initially, previous works like ReconNet \cite{ReconNet} directly apply the DNNs to CS restoration.
Nevertheless, they only exhibit similar restoration as BCS algorithm, and even worse than the later in some cases.
Subsequently, DR2Net remodels the refining networks as residual networks, namely $\hat{\boldsymbol{x}} = \boldsymbol{x}_P + \boldsymbol{x}_R$, and achieves  better CS restoration than ReconNet \cite{DR2, Deep_resiudal}. 
Meanwhile, some works combine traditional CS restoration and DNNs \cite{Learninginvert, LDAMP}. 
For instance, Learning Denoiser Approximate Message Passing (LDAMP) uses a traditional CS restoration method, i.e., Approximate Message Passing (AMP) \cite{GM-AMP, Yanting}, to obtain $\boldsymbol{x}_P$, and utilizes the refining networks to derive $\boldsymbol{x}_P$. 
In other words, LDAMP regards the refining networks as a powerful denoiser and integrates it into the traditional AMP method. 
It is noteworthy that LDAMP and DR2Net achieve better CS restoration than traditional BCS.
Fig. \ref{threarchitectures} shows the architecture of the three networks.

Though some DNNs like LDAMP outperforms traditional BCS algorithms, a problem of DNNs for CS restoration is that DNNs are vulnerable to perturbation \cite{CNNs_fooled, CNNs_fooled1, CNNs_robust}.
This problem implies that DNNs cannot precisely reconstruct input $\boldsymbol{x}$ from noisy measurement $\boldsymbol{y}$ \cite{DR2}.
More importantly, another fundamental problem of DNNs is that the working mechanism of DNNs is still not explainable and DNNs have been viewed as "black boxes" \cite{DNN_blackbox}. 
Therefore, we still don't have an explicit theory to explain how design an optimal DNNs for a specific application like CS restoration.

\section{A Statistical Framework of DNNs}

This section describes the proposed statistical framework of DNNs.
We demonstrate that the hidden layers of DNNs are equivalent to Gibbs distributions and the whole architecture of DNNs can be interpreted as a Bayesian hierarchical model.

To facilitate subsequent discussions, we assume that an arbitrary neural networks with $\boldsymbol{I}$ hidden layers denoted as $\{\boldsymbol{X; F_1; ...; F_I; \hat{Y}}\}$ is trained by $\mathcal{X}$ and $\mathcal{Y}$,  
where ${\textstyle \mathcal{X} = \{\boldsymbol{x}_n | \boldsymbol{x}_n \in \boldsymbol{R}^{W \times H}, n = 1, \cdots, N \}}$ denotes a training dataset 
and ${\textstyle \mathcal{Y} = \{\boldsymbol{y}_n | \boldsymbol{y}_n  \in \{0, 1\}^{L \times 1}, n = 1, \cdots, N \}}$ are training labels.
$\boldsymbol{X} \text{ and } \boldsymbol{Y}$ are the random variables related to $\mathcal{X}$ and $\mathcal{Y}$, respectively.
Similarly, $\boldsymbol{F_i}$ and $\boldsymbol{\hat{Y}}$ are the random variables for the $i$th hidden layer and output layer.

\subsection{A Statistical Representation of Hidden Layers}
\label{gibbs-dnns}

\textbf{Proposition 1:} \textit{The hidden layers of a neural network are equivalent to Gibbs distributions of the network input.} 

\begin{figure}[t]
\centering
\includegraphics[scale=0.35]{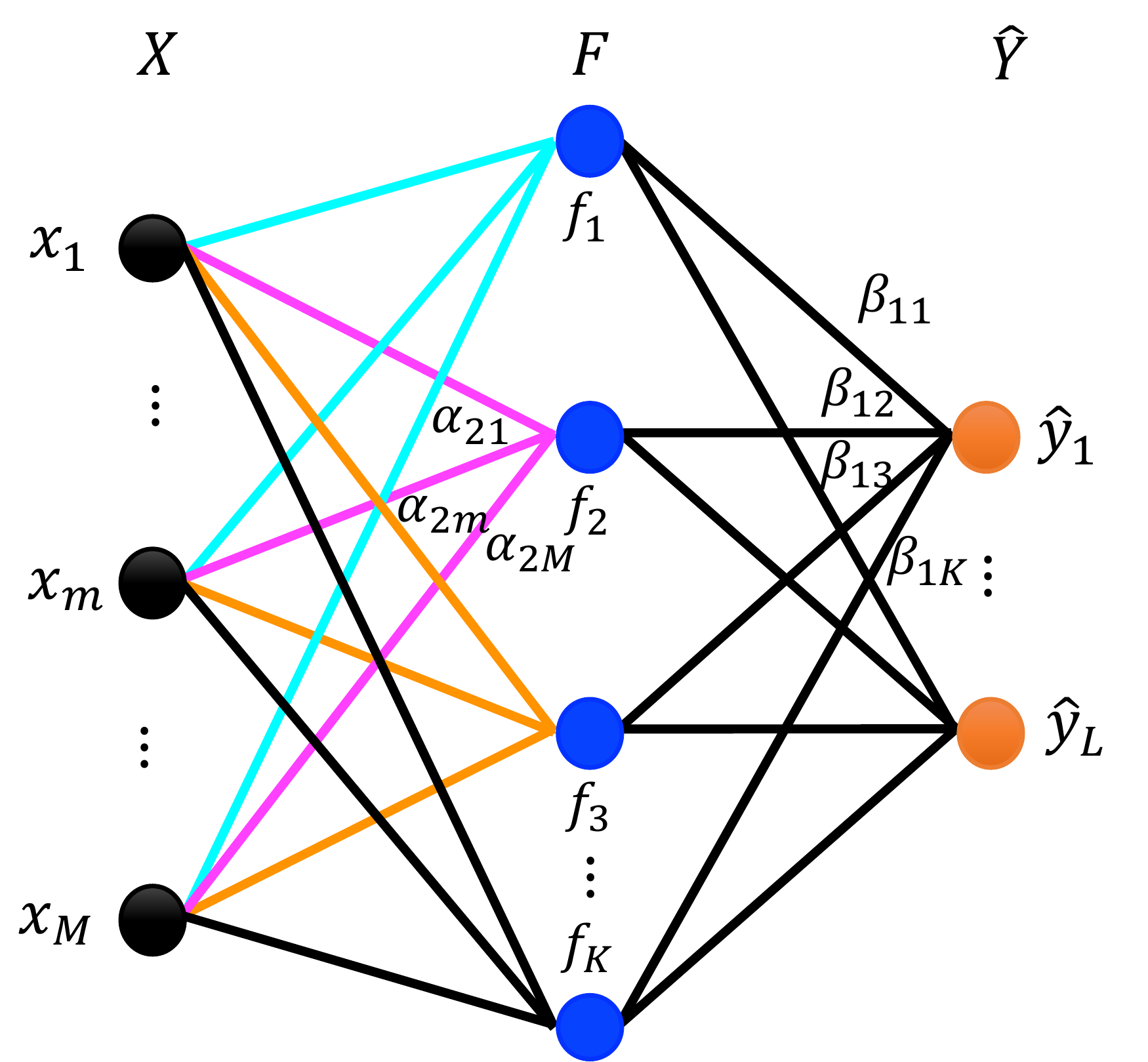}
\caption{\small{The architecture of a shallow neural network. In the hidden layer $\boldsymbol{F}$, each blue node is an activation function $f_k^{act}(\cdot)$, the lines with different colors indicate different linear filters $f_k(\cdot)$.}}	
\label{fig_ssn}
\end{figure}

Suppose there is a shallow neural network $\{\boldsymbol{X; F; \hat{Y}}\}$, in which $\boldsymbol{F}$ has $K$ neurons and $\boldsymbol{\hat{Y}}$ is softmax layer (Figure \ref{fig_ssn}), each output node $\hat{y}_l \in \boldsymbol{\hat{y}} = \{\hat{y}_1, ..., \hat{y}_L\}$ therefore can be formulated as
\begin{equation} 
\label{snn} 
{\textstyle
\hat{y}_l = \frac {1}{Z}\text{exp}\{\sum_{k=1}^{K}\beta_{lk} \cdot {f_k^{act}[f_k(\boldsymbol{x})]}\}
}
\end{equation}
where $\boldsymbol{x} \in \mathcal{X}$ is the input, $f_k^{act}(\cdot)$ is the activation function of the $k$th neuron, ${\textstyle f_k(\boldsymbol{x}) = (\sum_{m=1}^{M = W \times H}\alpha_{km} \cdot x_m)}$, and $\alpha_{km}$, $\beta_{lk}$ denote the weights. 
In addition, the partition function is defined as ${\scriptstyle Z = \sum_{l=1}^{L}\text{exp}\{\sum_{k=1}^{K}\beta_{lk} \cdot {f_k^{act}(f(\boldsymbol{x}))} \}}$.

We can obtain that $\boldsymbol{\hat{Y}}$ is a Gibbs distribution of $\mathcal{X}$ through comparing Equation (\ref{snn}) and (\ref{Gibbs}). 
Specifically, $\boldsymbol{\hat{Y}}$ assumes that $\mathcal{X}$ includes $L$ states $ \{\hat{y}_1, ..., \hat{y}_L\}$ and the \textit{energy} of each state is $g_l(\boldsymbol{x}) = \sum_{k=1}^{K}\beta_{lk} \cdot {f_k^{act}[f_k(\boldsymbol{x})]}$.
In particular, $g_l(\boldsymbol{x})$ is a linear combination of all neurons in $\boldsymbol{F}$.

We can also derive that $\boldsymbol{F}$ formulates a Gibbs distribution $p(\boldsymbol{F})$ of $\mathcal{X}$ based on Equation (\ref{Gibbs}).
Similarly, $\boldsymbol{F}$ assumes that $\mathcal{X}$ includes $K$ states $\{F_1, ..., F_K\}$ and the \textit{energy} of each state is  defined as $g_k(\boldsymbol{x}) = f_k^{act}[f_k(\boldsymbol{x})]$.
Therefore, $p(\boldsymbol{F})$ can be expressed as follows.
\begin{equation} 
\label{snn_f1} 
{\textstyle
p(F_{k}) = \frac {1}{Z_{\boldsymbol{F}}}\text{exp}\{f_k^{act}[f_k(\boldsymbol{x})]\}
}
\end{equation}

Overall, the hidden layer $\boldsymbol{F}$ defines the \textit{energy function} $g_k(\boldsymbol{x})$ of $p(F_{k})$.
Since the \textit{energy function} is a sufficient statistics of the Gibbs distribution $p(F_{k})$ \cite{it_book}, we can conclude that the hidden layers of a neural network are equivalent to Gibbs distributions of the network input.

Also note that the hidden layer defined in the network $\{\boldsymbol{X; F; \hat{Y}}\}$ is a fully connected layer, which assumes that there are finite $K$ states in $\mathcal{X}$ and uses the finite $K$ states to formulate a discrete Gibbs distribution of $\mathcal{X}$.
However, a discrete Gibbs distribution cannot explain the convolutional layer appropriately, since the output of a convolutional filter is a high-dimensional matrix, rather than a scalar to indicate the weight of a state. 
Alternatively, a convolutional layer with one or more non-linear linears can be formulated as a MRFs expressed as follows
\begin{equation} 
\label{cnn_mrf} 
{\textstyle
p(\boldsymbol{x}; \boldsymbol{\theta}) = \frac {1}{Z(\boldsymbol{\theta})}\text{exp}\{\sum_{k=1}^K[f_{k}^{NL}(f_k(\boldsymbol{x}); \boldsymbol{\theta}_k)]\}
}
\end{equation}
where $f_k(\cdot)$ is the convolutional filter and $f^{NL}(\cdot)$ represent non-linear layers, such as ReLU or Max pooling layer.

Similarly, we can prove that Proposition 1 holds for all hidden layers by redefining linear filter $f_k(\cdot)$ and non-linear function $f_{k}^{NL}(\cdot)$ based on Equation (\ref{snn_f1}) or Equation (\ref{cnn_mrf}).

\subsection{A Statistical Representation of DNNs}
\label{bhm-dnns}

\textbf{Proposition 2:} \textit{the whole architecture of DNNs can be interpreted as a Bayesian Hierarchical Model (BHM).}

Since the input of a hidden layer in DNNs is typically the output of its previous layer, it is more rational to state that a single hidden layer of DNNs defines a conditional Gibbs distribution given the previous layer based on Proposition 1.

Assuming there is a $\text{DNNs} = \{\boldsymbol{X; F_1; ...; F_I; \hat{Y}}\}$,
the distribution of $\boldsymbol{F_i}$ can be formulated a conditional Gibbs distribution $p(\boldsymbol{F_i|F_{i-1}})$ as Equation (\ref{snn_f1}) or (\ref{cnn_mrf}) depending on the functionality of $\boldsymbol{F_i}$.
Without loss of generality, $p(\boldsymbol{F_1|X})$ can be expressed as $p(\boldsymbol{F_1})$, since $\boldsymbol{X}$ will be replaced by $\mathcal{X}$, which is deterministic.

It has been proven that the $\text{DNNs} = \{\boldsymbol{X; F_1; ...; F_I; \hat{Y}}\}$ form a Markov chain expressed below \cite{DNN-Bottleneck, DNN-information},
\begin{equation} 
\label{mrf_dnns}
\boldsymbol{X \rightarrow F_1 \rightarrow \cdots \rightarrow F_I \rightarrow \hat{Y}}
\end{equation}
the distribution of DNNs thus can be formulated as follows. 
\begin{equation} 
\label{pdf_networks}
{\textstyle 
p(\boldsymbol{F_1; ...; F_I; \hat{Y}}) = p(\boldsymbol{F_1}) \cdot ...p(\boldsymbol{F_{i+1}|F_{i}}) \cdot ... p(\boldsymbol{\hat{Y}|F_{I}})
}
\end{equation}
The above distribution shows that the DNNs can be interpreted as a BHM with $\boldsymbol{I+1}$ levels, in which $\boldsymbol{F_i}$ formulates a conditional Gibbs distribution to process the features in $\boldsymbol{F_{i-1}}$ and serves as a prior distribution for the higher level $\boldsymbol{F_{i+1}}$.

Moreover, Bayes' theorem suggests that the above joint distribution of DNNs (Equation (\ref{pdf_networks})) can be decomposed into two components: a prior distribution $p(\boldsymbol{X})$ and a likelihood distribution $p(\boldsymbol{\hat{Y}|X})$ expressed as follows. 
\begin{equation} 
\label{bayesian_networks}
{\scriptstyle 
p(\boldsymbol{F_1; ...; F_I; \hat{Y}}) = \underbrace{{\scriptstyle p(\boldsymbol{F_1})  \cdot ... p(\boldsymbol{F_{i-1}|F_{i-2}})}}_{{\scriptstyle\text{prior}}} \underbrace{{\scriptstyle \cdot p(\boldsymbol{F_i|F_{i-1}}) \cdot ... p(\boldsymbol{\hat{Y}|F_{I}})}}_{{\scriptstyle \text{likelihood}}}
}
\end{equation}
That above equation suggests that DNNs utilize some hidden layers (i.e., $\boldsymbol{F_1} \cdots \boldsymbol{F_{i-1}}$) to learn $p(\boldsymbol{X})$ and the other layers (i.e., $\boldsymbol{F_i} \cdots \boldsymbol{\hat{Y}}$) are trained to learn $p(\boldsymbol{\hat{Y}|X})$ from $\mathcal{X}$ and $\mathcal{Y}$.

In summary, the proposed framework establishes an explicit correspondence between DNNs and statistical models in three aspects: 
(i) neuron defines the \textit{energy} of Gibbs distribution;
(ii) the hidden layers of DNNs formulate Gibbs distributions;
(iii) the whole architecture of DNNs can be interpreted as a BHM.
In particular, this statistical framework provides a novel Bayesian perspective to understand the architecture of DNNs.

\section{The insights into DNNs}

Based on the proposed statistical framework of DNNs, we specifies the application scope of DNNs, reveals a limitation of DNNs for CS restoration, establishes an overall picture of all existing BCS algorithms, and provides more explanation for the vulnerability of DNNs.

\subsection{The Application Scope of DNNs}

Based on Data Processing Inequality (DPI) \cite{it_book}, the Markov chain corresponding to the DNNs (Formula (\ref{mrf_dnns})) implies 
\begin{equation} 
H(\boldsymbol{X}) \geq I(\boldsymbol{X, F_1}) \cdots \geq I(\boldsymbol{X, \hat{Y}})
\end{equation}
where $H(\boldsymbol{X})$ is the entropy of $\boldsymbol{X}$ and $I(\boldsymbol{X, \hat{Y}})$ is the mutual information between $\boldsymbol{X}$ and $\boldsymbol{\hat{Y}}$.

The above inequality suggests that the information of DNNs input is great or equal to the output, which means that the application scope of DNNs is confine to the situation that $\mathcal{X}$ provides great or equal to the information of $\mathcal{Y}$.  
In terms of Bayesian hierarchical model, DNNs formulate the prior knowledge of $\mathcal{X}$ as a prior distribution $p(\boldsymbol{X})$ and establish an efficient representation of the output $p(\boldsymbol{\hat{Y}|X})$ based on the extracted prior knowledge.
Therefore, DNNs cannot learn an expressive Bayesian hierarchical model to derive the output unless the training dataset provides enough prior knowledge.

\subsection{A Limitation of DNNs for CS restoration}
\label{lim-dnns}

Assuming there is a $\text{DNNs} = \{\boldsymbol{Y; F_1; ...; F_I; \hat{X}}\}$ designed for CS restoration, where $\boldsymbol{Y}$ is the CS measurement and $\boldsymbol{\hat{X}}$ is the restored signal, 
We can derive that the DNNs formulate a Bayesian hierarchical model $p(\boldsymbol{\hat{X}, Y}) = p(\boldsymbol{Y})p(\boldsymbol{\hat{X}|Y})$.
However, the DNNs cannot extract much prior knowledge of $\boldsymbol{X}$ from $\boldsymbol{Y}$ since $\boldsymbol{Y}$ is compressed measurement.
Therefore, we cannot achieve the state-of-the-art performance through simply applying DNNs to CS restoration.

Since ReconNet is a specific DNNs for CS restoration ~~~(Fig. \ref{inverse_cnns_layers}), this limitation explains why it cannot outperform traditional BCS methods.
Moreover,  we can conclude that LDAMP achieves better restoration because it circumvents this limitation.
Specifically, the AMP algorithm derives the preliminary restoration $\boldsymbol{X}_p$ that provides much more prior knowledge of $\boldsymbol{X}$ than $\boldsymbol{Y}$ for LDAMP deriving $\boldsymbol{\hat{X}}$, which results in better restoration.

\subsection{A Reason for the Vulnerability of DNNs}

The vulnerability of DNNs implies that DNNs misclassify the input with small perturbations that are imperceptible to humans \cite{adversarial_training, adversarial_training2}.
The most widely accepted explanation for the vulnerability of DNNs is that the linear operators like convolutional filters magnify the perturbations, especially when the input is high-dimensional \cite{adversarial_training}.

\begin{figure}[t]
\centering
\includegraphics[scale=0.35]{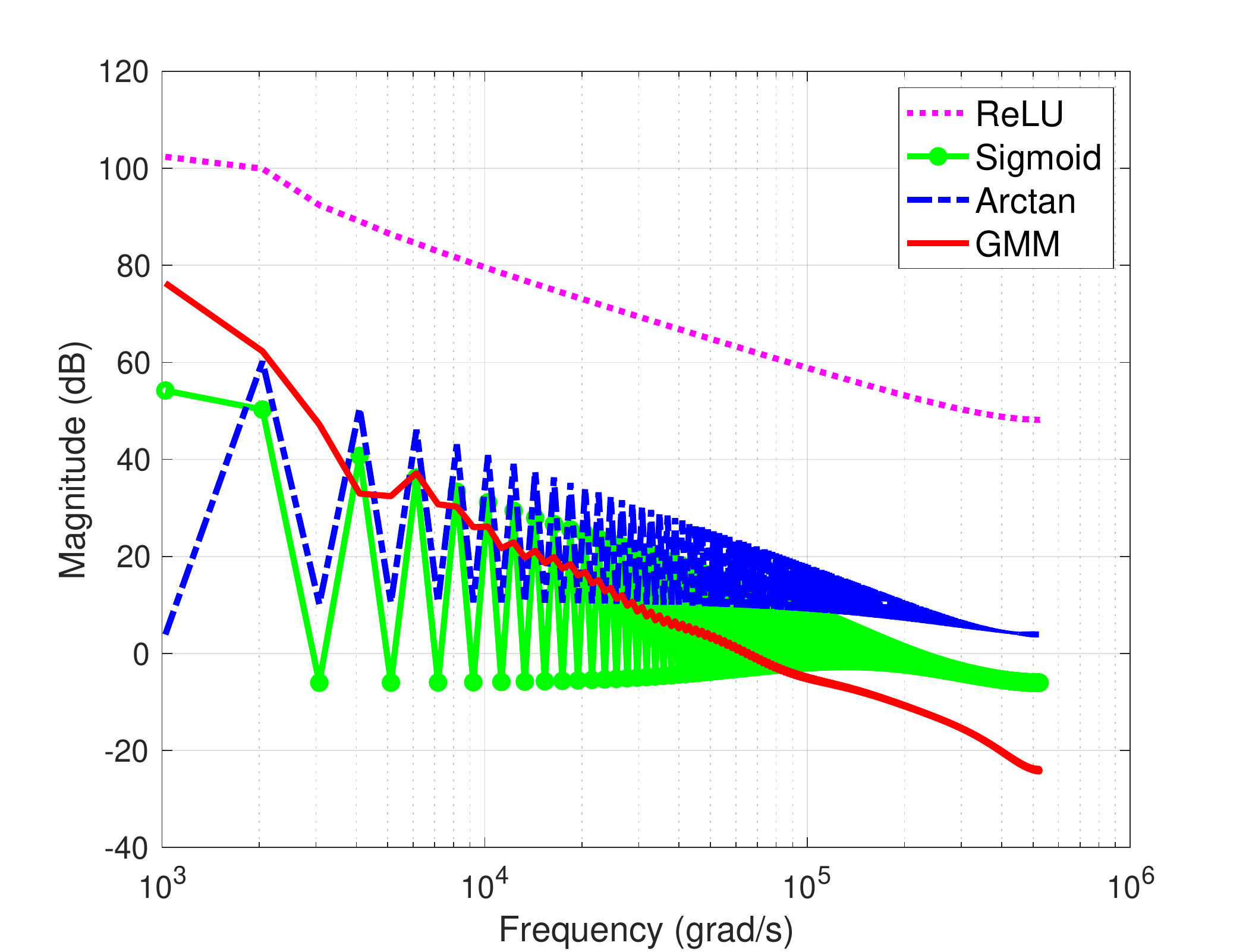}
\caption{\small{The magnitude spectrum of three popular activation functions, where Arctan indicates the arctangent function. As a comparison, GMM is short for Gaussian Mixture Model.
}}	
\label{act_freq}
\end{figure}

We discover another reason for the vulnerability of DNNs, i.e., the commonly used activation functions cannot denoise very well.
Fig. \ref{act_freq} shows the spectrum of three activation functions.
On the one hand, it shows that ReLU has the highest magnitude spectrum in all frequencies, hence ReLU can preserve the information of various features very well.
On the other hand, the spectrum implies that ReLU has the worst ability of denoising. 
In contrast, GMM, as a popular statistical model in image denoising field \cite{GSM-denoise, Yanting}, achieves a tradeoff. 
Fig. \ref{act_freq} shows that GMM has high magnitude spectrum at low frequencies and the lowest magnitude at high frequencies, which indicates that GMM can preserve feature information and implement denoising simultaneously.

Since the perturbation is typically viewed as noise, the vulnerability of DNNs has great effect on the performance of DNNs for CS restoration.
However, the most effective solution, namely adversarial training \cite{adversarial_training, adversarial_training1}, cannot improve the robustness of DNNs for CS restoration, since the variance of noise is unknown and varies in different CS applications such that we cannot train a DNNs for CS restoration in advance.
Therefore, designing new activation function is an alternative way to improve the robustness of DNNs for CS restoration. 

\subsection{An Overall Picture of BCS Algorithms}

We can divide all BCS algorithms into four categories depending on how to model the prior distribution $p(\boldsymbol{x})$ and likelihood distribution $p(\boldsymbol{y|x})$ in Table \ref{inverse_categories}, where $\boldsymbol{x}$ is the original signal and $\boldsymbol{y}$ is the CS measurement.
The first uses traditional distributions to model both $p(\boldsymbol{x})$ and $p(\boldsymbol{y|x})$, e.g., traditional BCS \cite{BCS}.
In contrast, the fourth category utilizes networks to estimate both $p(\boldsymbol{x})$ and $p(\boldsymbol{y|x})$, e.g., ReconNet \cite{ReconNet}.
The third category uses networks to estimate the prior knowledge of $\boldsymbol{x}$ and uses statistical methods to model the connection between $\boldsymbol{x}$ and $\boldsymbol{y}$, e.g., LDAMP \cite{LDAMP}.
To the best of our knowledge, there is no algorithm belongs to the second category. 
The categorization unifies traditional BCS methods and DNNs and provides an overall picture of BCS. 

\begin{table}[!b]
\centering
\caption{Four Categories of BCS Algorithms}
\label{inverse_categories}
\begin{tabu} to 0.43\textwidth { X[c]  X[c]  X[c]  X[c]}
 \hline
 \hline
 \textbf{Category} & \textbf{Prior} & \textbf{Likelihood} & \textbf{Example} \\
 \hline
 1  & statistics  & statistics  & BCS\cite{BCS} \\

 2  & statistics & networks  & {None} \\

 3  & networks  & statistics  & LDAMP\cite{LDAMP} \\

 4  & networks  & networks  & ReconNet\cite{ReconNet} \\
\hline
\end{tabu}
\end{table}

In summary, all existing BCS algorithms have limitations. 
The first cannot describe the prior knowledge of $\boldsymbol{x}$ precisely and comprehensively due to the limitation of traditional prior distributions discussed in Section \ref{lim_bcs}.
The fourth cannot achieve the state-of-the-art CS restoration because of the limitation of DNNs discussed in Section \ref{lim-dnns}.
Though the third circumvents the limitation of DNNs, the vulnerability of DNNs affects its performance in noisy situations.  

\section{Bayesian CNNs Method for CS Restoration}

This section describes a novel Bayesian CNNs (BCNNs) algorithm for CS restoration.
First, we propose a novel CNNs to model the prior distribution $p(\boldsymbol{x}; \boldsymbol{\theta})$ only.
Given the prior distribution $p(\boldsymbol{x}; \boldsymbol{\theta})$ corresponding to the CNNs, we derive the corresponding posterior distribution via the likelihood distribution of CS.
Finally, we propose a novel Bayesian inference algorithm based on auxiliary variable Gibbs sampler to infer the posterior distribution for CS restoration.

\subsection{A New Approach to Design DNNs for CS Restoration}

Unlike previous works designing an end-to-end DNNs to derive CS restoration directly, we propose a novel CNNs to model a prior distribution $p(\boldsymbol{x}; \boldsymbol{\theta})$ only, which brings three advantages over the existing DNNs for CS restoration:
(i) we can use high-dimensional dataset rather than CS measurement as the training dataset, which guarantees the proposed CNNs can extract enough prior knowledge to learn an expressive $p(\boldsymbol{x}; \boldsymbol{\theta})$;
(ii) we can decrease the complexity of networks, since the proposed CNNs only simulate the prior distribution $p(\boldsymbol{x}; \boldsymbol{\theta})$;
(iii) we can design new activation function to improve the robustness of the DNNs based on Proposition 1.

Given the $p(\boldsymbol{x}; \boldsymbol{\theta})$ generated from the proposed CNNs, we can derive the corresponding posterior distribution via the likelihood distribution of CS, and CS restoration can be achieved by inferring the posterior distribution.

\subsection{The Architecture of The Proposed CNNs}
\begin{figure}[t]
\centering
\centering\includegraphics[scale=0.15]{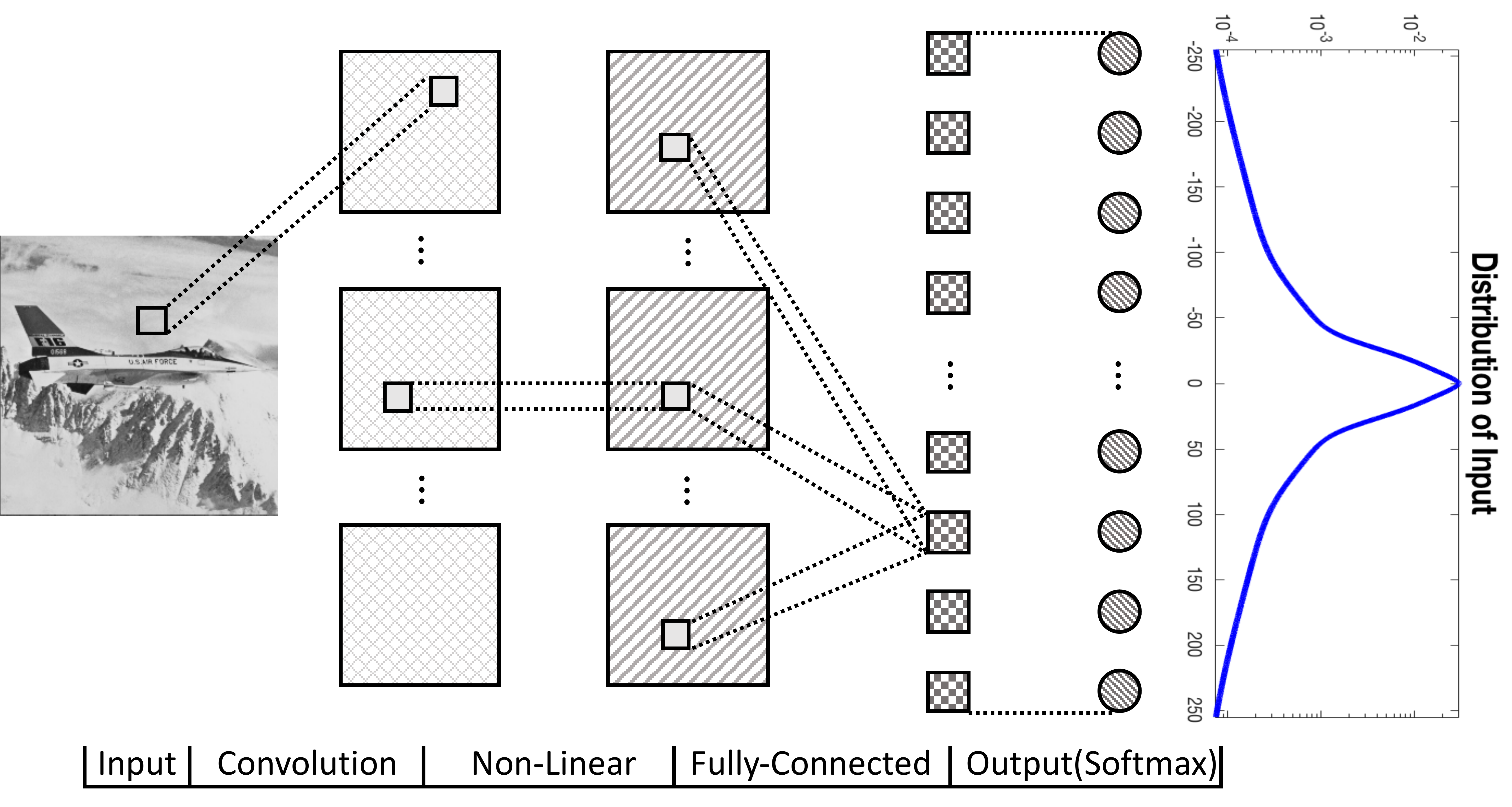}
\caption{The architecture of the proposed CNNs for modeling a prior distribution $p(\boldsymbol{x}; \boldsymbol{\theta})$ like the curve depicted in the right.}
\label{bcnn_architecture}
\end{figure}

The architecture of the proposed CNNs is depicted in Fig. \ref{bcnn_architecture}, which includes three hidden layers: convolutional, non-linear (i.e., activation function), and fully connected layers. 
The output layer is defined as softmax, and the corresponding prior distribution $p(\boldsymbol{x}; \boldsymbol{\theta})$ can be formulated as 
\begin{equation} 
\label{GCNN1}
{\scriptstyle
p(\boldsymbol{x}; \boldsymbol{\theta}) = \frac {1}{Z(\boldsymbol{\theta})} \text{exp}\{{\sum_{m=1}^{M}(f_m^{NL}(f_m(\boldsymbol{x})))\}}
}
\end{equation}
where $f_m(\cdot)$ and $f_m^{NL}(\cdot)$ are convolutional filter and activation function, respectively.
${\textstyle \sum_{m=1}^{M}(f_m^{NL}(f_m(\boldsymbol{x})))}$ represents the fully-connected layer.

Since GMM demonstrates better denoising ability than most popular activation functions, $f_m^{NL}(\cdot)$ is chosen as GMM to improve the robustness of the proposed CNNs.
It is formulated as ${\textstyle f^{NL}_m(\cdot) = \text{log}[\sum_{n=1}^{N}\pi_{mn} \cdot \mathcal{N}(f_m(\cdot); 0, {\sigma_b^2}/{\delta_n})]}$, where $\sigma_b^2$ is a fixed base variance, ${\textstyle {\boldsymbol{\delta} = \{\delta(1), \cdots, \delta(N)\}}}$ is a range of constant scales, and $\pi_{mn}$ denotes the weight of each Gaussian distribution. 
Therefore, the $p(\boldsymbol{x}; \boldsymbol{\theta})$ corresponding to the proposed CNNs can be formulated as
\begin{equation} 
\label{GCNN}
{\textstyle
p(\boldsymbol{x}; \boldsymbol{\theta}) = \frac {1}{Z(\boldsymbol{\theta})} \prod_{m=1}^{M}\sum_{n=1}^{N}\pi_{mn} \cdot \mathcal{N}(f_m(\boldsymbol{x}); 0, \frac{\sigma_b^2}{\delta(n)})
}
\end{equation}

Overall, the proposed CNNs formulates a prior distribution, in which the convolutional layer learn the features from the training dataset, and the activation function is defined as GMM to introduce non-linearity and implement denoising.

After specifying the CNNs architecture, the next problem is to construct the learning algorithm to optimize $\boldsymbol{\theta} = \{\boldsymbol{f, \pi}\}$, where $\boldsymbol{f} = [{f}_1, ..., {f}_m]$ and $\boldsymbol{\pi} = [\pi_{11}, ..., \pi_{NM}]$.
Here we choose Kullback-Leibler divergence (KLD) \cite{CD} as the criterion to measure the distance between ${\textstyle p(\boldsymbol{x}^{prior}; \boldsymbol{\theta})}$ and ${\textstyle p(\boldsymbol{x}^{data}; \boldsymbol{\theta})}$. KLD is defined as
\begin{equation} 
\label{KLD}
{\scriptstyle
\text{KLD}(p(\boldsymbol{x}^{prior}; \boldsymbol{\theta})||p(\boldsymbol{x}^{data}; \boldsymbol{\theta}))=\sum_{i}p(x^{data}(i); \boldsymbol{\theta}) \cdot \text{log}\frac{p(x^{data}(i); \boldsymbol{\theta})}{p(x^{prior}(i); \boldsymbol{\theta})}
}
 \end{equation}
where ${\textstyle p(\boldsymbol{x}^{prior}; \boldsymbol{\theta})}$ is the stationary distribution generated by sampling $p(\boldsymbol{x}; \boldsymbol{\theta})$, and ${\textstyle p(\boldsymbol{x}^{data}; \boldsymbol{\theta})}$ denotes the empirical distribution of the training dataset, namely the output of the proposed CNNs. 
In order to make $p(\boldsymbol{x}; \boldsymbol{\theta})$ model the statistical property of the training dataset as precisely as possible, the KLD should be minimized, which can be realized by gradient descent algorithm \cite{CD}. 

In practice, it is time-consuming to obtain ${\textstyle p(\boldsymbol{x}^{prior}; \boldsymbol{\theta})}$ by sampling $p(\boldsymbol{x}; \boldsymbol{\theta})$ continuously until stationary.
Therefore, Contrastive Divergence (CD) learning method \cite{CD1} is adopted to quickly optimize $\boldsymbol{\theta} = \{\boldsymbol{f, \pi}\}$, and it can be formulated as  
\begin{equation} 
\label{CD}
{\scriptstyle
\boldsymbol{\theta}_{n+1} = \boldsymbol{\theta}_{n} - \eta \cdot  [\langle \nabla CNN\rangle_{p(\boldsymbol{x}_{k}^{prior}; \boldsymbol{\theta}_n)} - \langle \nabla CNN\rangle_{p(\boldsymbol{x}^{data}; \boldsymbol{\theta}_n)}]
}
\end{equation}
where $\boldsymbol{\theta}_n$ denote the optimized parameters at the $n$th training epoch,
${\scriptstyle \nabla CNN = \frac{\partial \text{log}(p(\boldsymbol{x}, \boldsymbol{\theta})}{\partial \boldsymbol{\theta}}}$, $\langle \rangle_{p}$ indicates the average over $p$, and $\eta \in (0,1]$ denotes the learning rate. 

Since CD learning algorithm only takes few $k$ sampling iterations to estimate $p(\boldsymbol{x}; \boldsymbol{\theta})$, it can increase training speed greatly while guarantee similar training result.
In summary, the learning algorithm for the proposed CNNs is presented in Algorithm \ref{bcnn_prior}. 

\begin{algorithm}[t!]
  \caption{CNNs learning algorithm for \textit{prior} distribution}  
  \label{bcnn_prior}  
  \begin{algorithmic}[1]  
    \Require
     training data $\boldsymbol{TD}$, CNNs model 
    \begingroup
    \everymath{\footnotesize}
    \small
    \State
    \textbf{initialize} 
    \State
    \hspace{0.4cm} GSM scales ${\delta_n}$, base variance $\sigma_b^2$
    \State
    \hspace{0.4cm} CNNs parameters $\boldsymbol{\theta} = \{\boldsymbol{f, \pi}\}$
    \State
    \hspace{0.4cm} sampling iteration $k$
    \State
    \hspace{0.4cm} training iteration count $n = 0$
    \State
    \hspace{0.4cm} training parameters, e.g., learning rate $\eta$, batch size.
    
    \Repeat
      \State prepare training batch $\boldsymbol{TB}_n$ from $\boldsymbol{TD}$
      \State compute  $\langle \nabla CNN\rangle_{p(\boldsymbol{x}^{data}; \boldsymbol{\theta}_n)}$ based on $\boldsymbol{TB}_n$
      \State obtain training label $\boldsymbol{TL}_n$ via sampling $ p(\boldsymbol{x}^{prior}_k; \boldsymbol{\theta}_n)$
      \State compute $\langle \nabla CNN\rangle_{p(\boldsymbol{x}_{k}^{prior}; \boldsymbol{\theta}_n)}$ based on $\boldsymbol{TL}_n$
      \State update $\boldsymbol{\theta}_{n}$ based on gradient descent method (\ref{CD})
      \State n $\leftarrow$ n + 1
    \Until{($|\boldsymbol{\theta_{n+1}} - \boldsymbol{\theta_{n}}| < \xi$ or $n > N$)}     
    
    \endgroup
    \Ensure  
      optimal parameter $\boldsymbol{\theta^{*}} = \boldsymbol{\theta_{n}}$  
  \end{algorithmic}  
\end{algorithm}

\subsection{Bayesian CNNs Inference for CS restoration}

In the framework of traditional BCS, we need derive the posterior distribution $p(\boldsymbol{x}|\boldsymbol{A}, \boldsymbol{y}, \boldsymbol{z}, \sigma_n^2; \boldsymbol{\theta})$ based on the prior distribution $p(\boldsymbol{x}$; $\boldsymbol{\theta})$ and the likelihood distribution $p(\boldsymbol{y}|\boldsymbol{x}, \boldsymbol{A}, \sigma_n^2)$ for CS restoration. 
However, the $p(\boldsymbol{x}$; $\boldsymbol{\theta})$ is intractable for most conventional inference algorithms \cite{Geman}. 
Alternatively, we utilize auxiliary variable Gibbs sampler to simplify the $p(\boldsymbol{x}$; $\boldsymbol{\theta})$ for deriving an applicable posterior inference algorithm \cite{xinjie, half-quadratic}. 

Since GMM can be viewed as a discrete case of GSM model \cite{GSM-stat}, which is defined as
\begin{equation} 
\label{GSM}
{\scriptstyle
p(\boldsymbol{x}, \boldsymbol{z}) = \int_{-\infty}^{\infty} \frac{1}{2\pi|z^2\boldsymbol{\Sigma}|^{1/2}} \text{exp}(-\frac{\boldsymbol{X^{T}\Sigma^{-1}X})}{2z^2}) \cdot p(z)dz
}
\end{equation}
where $\boldsymbol{z}$ denotes an auxiliary random vector $\boldsymbol{z} \in \{1, ..., N\}^M$ to represent the scale $\delta_n$, 
the $p(\boldsymbol{x}$; $\boldsymbol{\theta})$ can be augmented into a joint distribution $p(\boldsymbol{x, z; \theta})$.

Furthermore, two conditional distributions $p(\boldsymbol{x|z; \theta})$ and $p(\boldsymbol{z|x; \theta})$ corresponding to $p(\boldsymbol{x, z; \theta})$ can be derived below.
\begin{equation} 
{\scriptstyle
p(z_{mn}|\boldsymbol{x; \theta}) \propto \pi_{mn} \cdot \mathcal{N}(\boldsymbol{f}_m(\boldsymbol{x}); 0, \frac{\sigma_b^2}{\delta_n})
}
\end{equation}
\begin{equation} 
\label{x_given_z}
{\scriptstyle
p(\boldsymbol{x|z; \theta}) \propto \prod_{m=1}^{M} \mathcal{N}(\boldsymbol{f}_m(\boldsymbol{x}); 0, \frac{\sigma_b^2}{\boldsymbol{z}_m})
}
\end{equation}
Ultimately, the posterior distribution $p(\boldsymbol{x}|\boldsymbol{A}, \boldsymbol{y}, \boldsymbol{z}, \sigma_n^2; \boldsymbol{\theta})$ of CS can be formulated below based on (\ref{x_given_z}) and (\ref{likelihood}).  
\begin{equation} 
{\scriptstyle
p(\boldsymbol{x}|\boldsymbol{A}, \boldsymbol{y}, \boldsymbol{z}, \sigma_n^2; \boldsymbol{\theta}) \propto \prod_{m=1}^{M} \mathcal{N}(\boldsymbol{f}_m(\boldsymbol{x}); 0, \frac{\sigma_b^2}{\boldsymbol{z}_m}) \cdot p(\boldsymbol{y}|\boldsymbol{x}, \boldsymbol{A}, \sigma_n^2)
}
\end{equation}
Since all $\boldsymbol{f}_m$ and $\boldsymbol{A}$ are linear, $p(\boldsymbol{x}|\boldsymbol{A}, \boldsymbol{y}, \boldsymbol{z}, \sigma_n^2; \boldsymbol{\theta})$ can be rewritten in matrix form to get more intuitive expression.
\begin{equation} 
\label{GSM-posterior}
{\scriptstyle
p(\boldsymbol{x}|\boldsymbol{A}, \boldsymbol{y}, \boldsymbol{z}, \sigma_n^2; \boldsymbol{\theta}) \propto \text{exp}\{-({\boldsymbol{F}}\boldsymbol{x}-\boldsymbol{{\mu}})^T\boldsymbol{{\Sigma}}^{-1}({\boldsymbol{F}}\boldsymbol{x}-\boldsymbol{{\mu}})\} 
}
\end{equation}
where 
$$
{\scriptstyle
\boldsymbol{{F}} = 
\begin{bmatrix} 
{\scriptstyle \boldsymbol{f}_1 }\\
{\scriptstyle \vdots } \\
{\scriptstyle \boldsymbol{f}_m }\\
{\scriptstyle \boldsymbol{A} }
\end{bmatrix};
\boldsymbol{{\Sigma}} = 
\begin{bmatrix} 
{\scriptstyle \frac{\sigma_b^2}{\boldsymbol{z}_1} \boldsymbol{I}} & \dots & {\scriptstyle \boldsymbol{0}} & {\scriptstyle \boldsymbol{0}} \\
\vdots & \ddots & {\scriptstyle \boldsymbol{0}} & \vdots \\ 
{\scriptstyle \boldsymbol{0}} & \dots & {\scriptstyle \frac{\sigma_b^2}{\boldsymbol{z}_M} \boldsymbol{I}} & \boldsymbol{0} \\
{\scriptstyle \boldsymbol{0}} & \dots & {\scriptstyle \boldsymbol{0}} & {\scriptstyle{\sigma_n^2} \boldsymbol{I}}
\end{bmatrix};
\boldsymbol{{\mu}} = 
\begin{bmatrix} 
{\scriptstyle \boldsymbol{0} }\\
{\scriptstyle \vdots } \\
{\scriptstyle \boldsymbol{0} }\\
{\scriptstyle \boldsymbol{y} }
\end{bmatrix} 
}
$$

We can find that Equation (\ref{GSM-posterior}) is a Gaussian distribution, such that CS restoration can be inferred through alternately sampling $p(\boldsymbol{x}|\boldsymbol{A}, \boldsymbol{y}, \boldsymbol{z}, \sigma_n^2; \boldsymbol{\theta})$ and $p(\boldsymbol{z|x; \theta})$ in Algorithm \ref{bcnn_infer}.

\section{Experimental Results}

This section presents our experimental results.
First, BCNNs are evaluated in CS image restoration field given different hyper-parameters.
Second, BCNNs are compared to the state-of-the-art CS restoration algorithms in noiseless and noisy situations.
BCNNs are implemented by MATLAB. All related simulation code, training dataset are available online\footnote{Available at \url{https://github.com/EthanLan/BCNN}}.

\begin{algorithm}[t!]  
  \caption{Bayesian CNNs algorithm for CS restoration}  
   \label{bcnn_infer}  
  \begin{algorithmic}[1]  
    \Require
     $\boldsymbol{y}$, $\boldsymbol{A}$, $p(\boldsymbol{x}$; $\boldsymbol{\theta})$, and $p(\boldsymbol{y}|\boldsymbol{x}, \boldsymbol{A}, \sigma_n^2)$
    \begingroup
    \everymath{\footnotesize}
    \small
    \State
    \textbf{initialize} 
    \State
    \hspace{0.4cm} randomly initialize $\boldsymbol{x}$, $\boldsymbol{z}$, $\sigma_n^2$
    \State
    \hspace{0.4cm} sampling iteration $n$
    
    \Repeat
      \State sampling $p(z_{mn}|\boldsymbol{x; \theta})$
      \State sampling $p(\boldsymbol{x}|\boldsymbol{A}, \boldsymbol{y}, \boldsymbol{z}, \sigma_n^2; \boldsymbol{\theta})$     
      \State sampling $p(\sigma_n^{-2}|\boldsymbol{x},\boldsymbol{y}, \boldsymbol{A}) = \text{Gamma}(\frac{M}{2}+1,\frac{2}{\Vert \boldsymbol{y} - \boldsymbol{Ax} \Vert^2})$
      \State $n \leftarrow n-1$  
       
    \Until{($n = 0$)}     
    
    \endgroup
    \Ensure  
      recovery signal $\boldsymbol{x^{*}} = \boldsymbol{x}$    
  \end{algorithmic}  
\end{algorithm}

\subsection{Experimental Setup}

There are three hyper-parameters of BCNNs: convolutional depth $F_n$ (i.e., the number of convolutional filters $f_m$), filter dimension $\mathcal{N}_d$, and GSM scales $\boldsymbol{\delta}$ need to initialize.

To evaluate the influence of different hyper-parameters on BCNNs, we instantiate five models of BCNNs, and they are summarized in Table \ref{BCNN_model}.
For example, BCNN2 has four convolutional filters, i.e., $F_n = 4$, and the dimension of each filter is defined as $\mathcal{N}_2$. 
The scales $\boldsymbol{\delta}$ of BCNN2 are initialized as ${\textstyle \boldsymbol{\delta}_1}$. 
Therefore, BCNN2 can be formulated as follows. 
\begin{equation} 
\label{BCNN2}
{\scriptstyle
p(\boldsymbol{x}; \boldsymbol{\theta}) = \frac {1}{Z(\boldsymbol{\theta})} \prod_{m=1}^{4}\sum_{n=1}^{5}\pi_{mn} \cdot \mathcal{N}(f_m(\boldsymbol{x}); 0, \frac{\sigma_b^2}{{\delta_{1}}(n)})
}
\end{equation}

We can derive that BCNN2 has 56 parameters need to learn, i.e., $|\boldsymbol{\theta}| = F_n \times |f_m| + |\boldsymbol{\pi}| = 56$, where $|f_m|$ and $|\boldsymbol{\pi}|$ denote the number of parameters of $f_m$ and $\pi_{mn}$, respectively.
Table \ref{BCNN_model} shows that larger $\mathcal{N}_d$, $F_n$, or $\boldsymbol{\delta}$ requires more parameters, hence $|\boldsymbol{\theta}|$ can indicate the complexity of each model. 

\begin{table}[htp]
  \centering
  \caption{BCNNs Models with Different Hyper-parameters}
  \begin{threeparttable}
  \begin{tabular*} {0.4\textwidth}{@{\extracolsep{\fill}}  c c c c c}
    \hline
    \hline
    {\footnotesize Model} & {\footnotesize $\mathcal{N}_d$} & {\footnotesize $F_n$} & {\footnotesize $\boldsymbol{\delta}$} & {\footnotesize $|\boldsymbol{\theta}|$}
    \\ 
    \hline
    {\footnotesize BCNN1} & {\footnotesize $\mathcal{N}_1$} & {\footnotesize 4} & {\footnotesize $\boldsymbol{\delta}_1$} & {\footnotesize 40}
    \\ 
    {\footnotesize BCNN2} & {\footnotesize $\mathcal{N}_2$} & {\footnotesize 4} & {\footnotesize $\boldsymbol{\delta}_1$} & {\footnotesize 56}
    \\ 
    {\footnotesize BCNN3} & {\footnotesize $\mathcal{N}_2$} & {\footnotesize 8} & {\footnotesize $\boldsymbol{\delta}_1$} & {\footnotesize 113}
    \\ 
    {\footnotesize BCNN4} & {\footnotesize $\mathcal{N}_2$} & {\footnotesize 8} & {\footnotesize $\boldsymbol{\delta}_2$} & {\footnotesize 136}
    \\ 
    {\footnotesize BCNN5} & {\footnotesize $\mathcal{N}_3$} & {\footnotesize 24} & {\footnotesize $\boldsymbol{\delta}_2$} & {\footnotesize 792}
    \\
    \hline
  \end{tabular*}
  \begin{tablenotes}
            \item $\mathcal{N}_1$ means a filter only includes the nearest four neighbors.
            \item $\mathcal{N}_2$ is $3 \times 3$ filter, and $\mathcal{N}_3$ is $5 \times 5$ filter.
            \item ${\textstyle \boldsymbol{\delta}_1 = \{\text{exp}(-7,-3,0,3,7)\}}$, ${\textstyle \boldsymbol{\delta}_2 = \{\text{exp}(\pm7,\pm5,\pm3, \pm1)\}}$.
        \end{tablenotes}
  \end{threeparttable}
  \label{BCNN_model}
\end{table}

Six recently developed CS restoration methods: 
LASSO\footnote{Available at \url{http://sparselab.stanford.edu/}}, 
SSM\footnote{Available at \url{https://sites.google.com/site/link2yulei/publications}}, 
TV\footnote{Available at \url{http://www.caam.rice.edu/~optimization/L1/TVAL3/}},
ReconNet\footnote{Available at \url{https://github.com/KuldeepKulkarni/ReconNet}}, 
DR2Net\footnote{Available at \url{https://github.com/coldrainyht/caffe_dr2/tree/master/DR2}}, 
and LDAMP\footnote{Available at \url{https://github.com/ricedsp/D-AMP_Toolbox}} are chosen as references.
LASSO is a well-known CS restoration algorithm \cite{Lasso}. 
SSM is a BCS method integrating the cluster structure in $\mathcal{N}_1$ dimension by formulating the prior distribution as the Spike and Slab Model (SSM) defined below \cite{BCS-neig}. 
\begin{equation} 
{\textstyle 
p(\boldsymbol{x}) = (1 - \pi)\delta_0 + \pi \mathcal{N}(\boldsymbol{x}; 0, \sigma^2)
}
\end{equation}
where $\delta_0$ is the Dirac function centered at zero, and $\pi$ is commonly extended to a linear filter to describe the cluster structure. 
As a counterpart, TV is a deterministic CS method to include the same signal structure as SMM \cite{TVAL3}.

ReconNet, DR2Net and LDAMP are three DNNs designed for CS restoration. 
For a fair comparison, the refining networks of DR2Net and LDAMP are redesigned to keep the same as ReconNet except for some necessary layers for the residual technique.
In other words, the refining networks of the three DNNs have six convolutional layers with the same convolutional filters and six ReLU layers depicted in Fig. \ref{inverse_cnns_layers}.

\begin{table*}[h!]
  \centering
   \caption{BCNNs Restoration Performance [PSNR/SSIM] Given Different Hyper-parameters}
  \begin{tabular}{ c c c c c c c}
    {\footnotesize Original} & {\footnotesize SSM} &{\footnotesize BCNN1} & {\footnotesize BCNN2} & {\footnotesize BCNN3} & {\footnotesize BCNN4} & {\footnotesize BCNN5}   
    \\ 
    \begin{minipage}{.12\textwidth}
      \includegraphics[width=\textwidth]{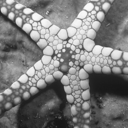}
    \end{minipage}
    &
      \begin{minipage}{.12\textwidth}
      \includegraphics[width=\textwidth]{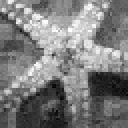}
     \end{minipage}
    &
      \begin{minipage}{.12\textwidth}
      \includegraphics[width=\textwidth]{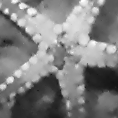}
     \end{minipage}
    & 
      \begin{minipage}{.12\textwidth}
      \includegraphics[width=\textwidth]{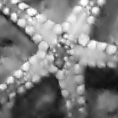}
     \end{minipage}
    &
    \begin{minipage}{.12\textwidth}
      \includegraphics[width=\textwidth]{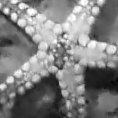}
     \end{minipage}
     &
    \begin{minipage}{.12\textwidth}
      \includegraphics[width=\textwidth]{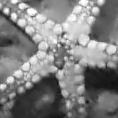}
     \end{minipage}
     &
    \begin{minipage}{.12\textwidth}
      \includegraphics[width=\linewidth]{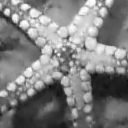}
     \end{minipage}
    \\ 
    {\footnotesize Starfish} & {\footnotesize $21.28\text{dB}/0.58$} & {\footnotesize $22.42\text{dB}/0.64$} & {\footnotesize $23.98\text{dB}/0.74$} & {\footnotesize $24.37\text{dB}/0.76$} & {\footnotesize $24.45\text{dB}/0.77$} & {\footnotesize $\boldsymbol{24.56\text{dB}/0.77}$} 
    \\ 
    \begin{minipage}{.12\textwidth}
      \includegraphics[width=\textwidth]{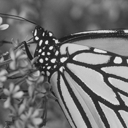}
    \end{minipage}
    &
      \begin{minipage}{.12\textwidth}
      \includegraphics[width=\textwidth]{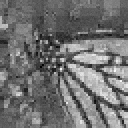}
     \end{minipage}
    &
      \begin{minipage}{.12\textwidth}
      \includegraphics[width=\textwidth]{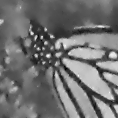}
     \end{minipage}
    & 
      \begin{minipage}{.12\textwidth}
      \includegraphics[width=\textwidth]{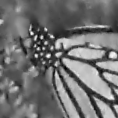}
     \end{minipage}
    &
    \begin{minipage}{.12\textwidth}
      \includegraphics[width=\textwidth]{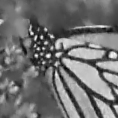}
     \end{minipage}
     &
    \begin{minipage}{.12\textwidth}
      \includegraphics[width=\textwidth]{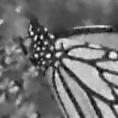}
     \end{minipage}
     &
    \begin{minipage}{.12\textwidth}
      \includegraphics[width=\linewidth]{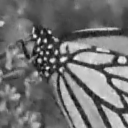}
     \end{minipage}
    \\ 
    {\footnotesize Butterfly} & {\footnotesize $19.54\text{dB}/0.58$} & {\footnotesize $22.61\text{dB}/0.75$} &{\footnotesize $24.05\text{dB}/0.81$} & {\footnotesize $24.40\text{dB}/0.82$} & {\footnotesize $24.59\text{dB}/0.82$} & {\footnotesize $\boldsymbol{24.80\text{dB}/0.84}$} 
    \\
     \begin{minipage}{.12\textwidth}
      \includegraphics[width=\textwidth]{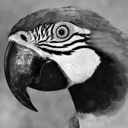}
    \end{minipage}
    &
      \begin{minipage}{.12\textwidth}
      \includegraphics[width=\textwidth]{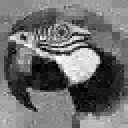}
     \end{minipage}
    &
      \begin{minipage}{.12\textwidth}
      \includegraphics[width=\textwidth]{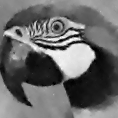}
     \end{minipage}
    & 
      \begin{minipage}{.12\textwidth}
      \includegraphics[width=\textwidth]{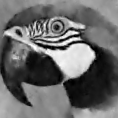}
     \end{minipage}
    &
    \begin{minipage}{.12\textwidth}
      \includegraphics[width=\textwidth]{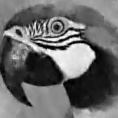}
     \end{minipage}
     &
    \begin{minipage}{.12\textwidth}
      \includegraphics[width=\textwidth]{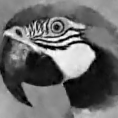}
     \end{minipage}
     &
    \begin{minipage}{.12\textwidth}
      \includegraphics[width=\linewidth]{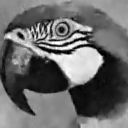}
     \end{minipage}
    \\ 
    {\footnotesize Parrot} & {\footnotesize $20.87\text{dB}/0.56$} & {\footnotesize $25.32\text{dB}/0.82$} & {\footnotesize $25.92\text{dB}/0.84$} & {\footnotesize $26.53\text{dB}/0.85$} & {\footnotesize $\boldsymbol{26.72\text{dB}/0.85}$} & {\footnotesize $26.61\text{dB}/0.84$} 
    \\
  \end{tabular}
  \label{BCNN_self}
\end{table*}

The same training dataset of ReconNet and DR2Net are used to train BCNNs and LDAMP, which consists of 91 natural images. 
We uniformly extract 21,668 $20 \times 20$ image patches from this dataset for training. 
Peak Signal-to-Noise Rate (PSNR), Structural Similarity Index (SSIM), and KLD are chosen as the quantitative criteria.

\subsection{Evaluation of BCNNs for CS restoration}

\begin{figure}[!b]
\centering
\centering\includegraphics[scale=0.47]{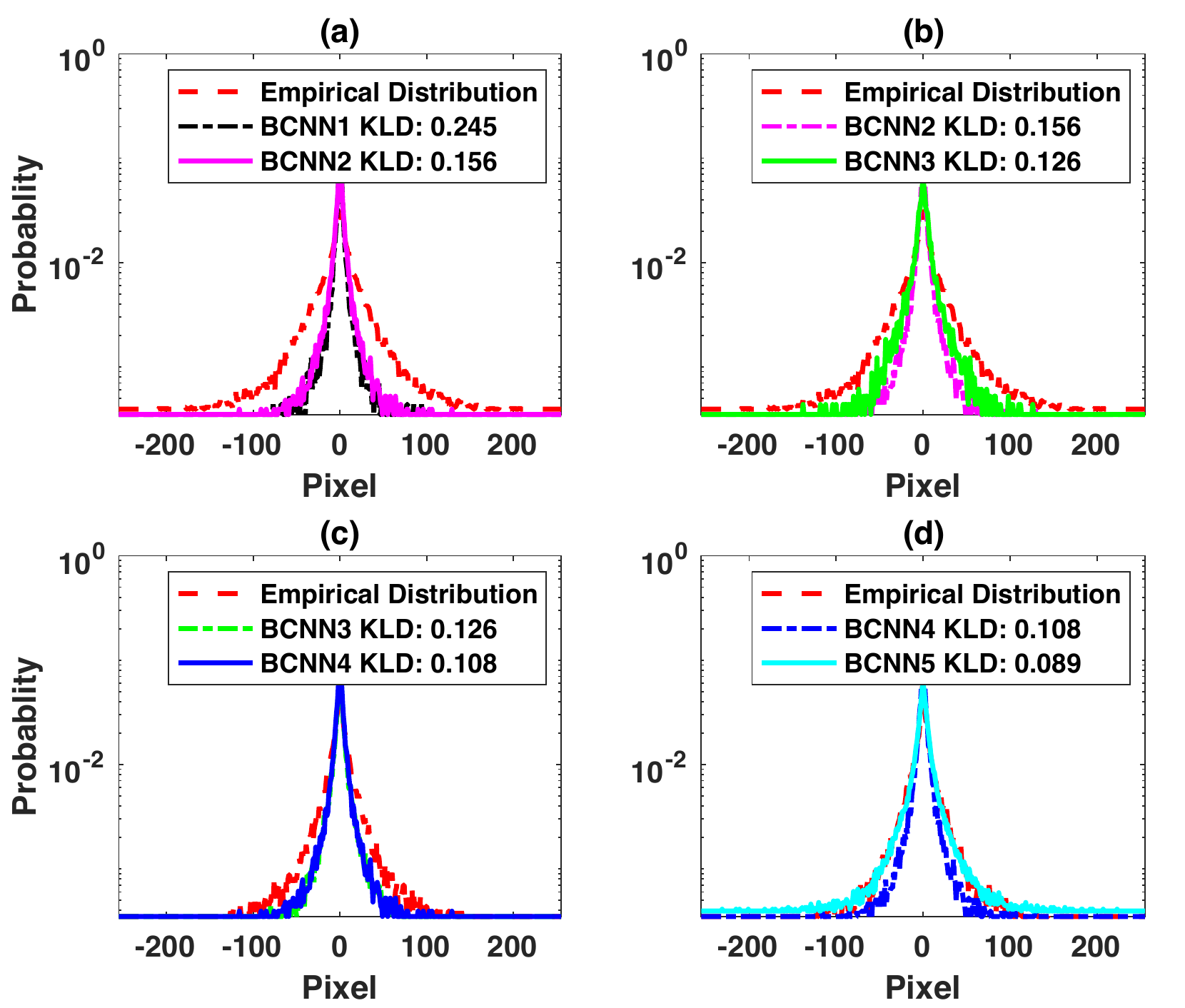}
\caption{(a) the influence of increasing $\mathcal{N}_1$ to $\mathcal{N}_2$, (b) the influence of increasing $F_4$ to $F_8$, (c) the influence of increasing $\boldsymbol{\delta}_1$ to $\boldsymbol{\delta}_2$, (d) the influence of increasing $\mathcal{N}_d$ and $F_n$ both.}
\label{kld_bcnns}
\end{figure}

Since the $p(\boldsymbol{x}; \boldsymbol{\theta})$ plays a vital role in CS restoration, we first evaluate the ability of BCNNs to model $p(\boldsymbol{x}; \boldsymbol{\theta})$ given different hyper-parameters.
We compare the KLD between an empirical distribution and the samples of five BCNN models.
The empirical distribution (the red curve in Fig. \ref{kld_bcnns}) is generated from a testing dataset including 20 image patches randomly chosen from the training dataset.
The samples of five BCNN models are generated by sampling the $p(\boldsymbol{x}; \boldsymbol{\theta})$.

\begin{table*}[h!]
  \centering
   \caption{Restoration Comparison [PSNR/SSIM] in Noiseless Situation}
  \begin{tabular}{ c c c c c c c}
    {\footnotesize Original} & {\footnotesize SSM} & {\footnotesize TV} & {\footnotesize ReconNet} & {\footnotesize DR2Net} & {\footnotesize LDAMP} & {\footnotesize BCNNs}
    \\ 
    \begin{minipage}{.12\textwidth}
      \includegraphics[width=\textwidth]{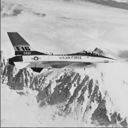}
    \end{minipage}
    &
      \begin{minipage}{.12\textwidth}
      \includegraphics[width=\textwidth]{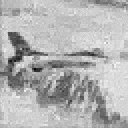}
     \end{minipage}
    & 
      \begin{minipage}{.12\textwidth}
      \includegraphics[width=\textwidth]{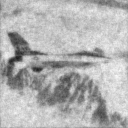}
     \end{minipage}
    &
    \begin{minipage}{.12\textwidth}
      \includegraphics[width=\textwidth]{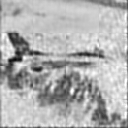}
     \end{minipage}
     &
    \begin{minipage}{.12\textwidth}
      \includegraphics[width=\textwidth]{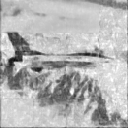}
     \end{minipage}
     &
    \begin{minipage}{.12\textwidth}
      \includegraphics[width=\linewidth]{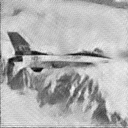}
     \end{minipage}
     &
    \begin{minipage}{.12\textwidth}
      \includegraphics[width=\linewidth]{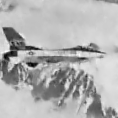}
     \end{minipage}
    \\   
    {\footnotesize Plane} & {\footnotesize $22.06\text{dB}/0.59$}  & {\footnotesize $23.19\text{dB}/0.62$} & {\footnotesize $22.23\text{dB}/0.65$}  & {\footnotesize $23.89\text{dB}/0.70$}  & {\footnotesize 	$27.23\text{dB}/0.80$}  & {\footnotesize $\boldsymbol{27.47\text{dB}/0.86}$} 
    \\ 
    \begin{minipage}{.12\textwidth}
      \includegraphics[width=\textwidth]{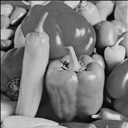}
    \end{minipage}
    &
      \begin{minipage}{.12\textwidth}
      \includegraphics[width=\textwidth]{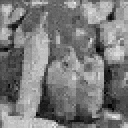}
     \end{minipage}
    & 
      \begin{minipage}{.12\textwidth}
      \includegraphics[width=\textwidth]{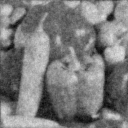}
     \end{minipage}
    &
    \begin{minipage}{.12\textwidth}
      \includegraphics[width=\textwidth]{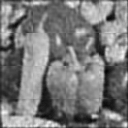}
     \end{minipage}
     &
    \begin{minipage}{.12\textwidth}
      \includegraphics[width=\textwidth]{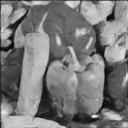}
     \end{minipage}
     &
    \begin{minipage}{.12\textwidth}
      \includegraphics[width=\linewidth]{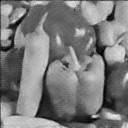}
     \end{minipage}
     &
    \begin{minipage}{.12\textwidth}
      \includegraphics[width=\linewidth]{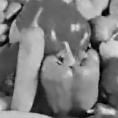}
     \end{minipage}
    \\   
    {\footnotesize Peppers} & {\footnotesize $21.79\text{dB}/0.61$}  & {\footnotesize $23.45\text{dB}/0.71$} & {\footnotesize $23.00\text{dB}/0.73$}  & {\footnotesize $25.17\text{dB}/0.78$}  & {\footnotesize 	$29.01\text{dB}/0.90$}  & {\footnotesize $\boldsymbol{29.73\text{dB}/0.90}$} 
    \\ 
    \\ 
    \begin{minipage}{.12\textwidth}
      \includegraphics[width=\textwidth]{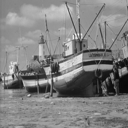}
    \end{minipage}
    &
      \begin{minipage}{.12\textwidth}
      \includegraphics[width=\textwidth]{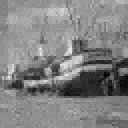}
     \end{minipage}
    & 
      \begin{minipage}{.12\textwidth}
      \includegraphics[width=\textwidth]{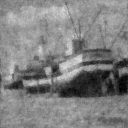}
     \end{minipage}
    &
    \begin{minipage}{.12\textwidth}
      \includegraphics[width=\textwidth]{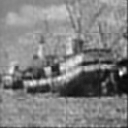}
     \end{minipage}
     &
    \begin{minipage}{.12\textwidth}
      \includegraphics[width=\textwidth]{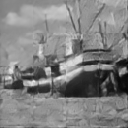}
     \end{minipage}
     &
    \begin{minipage}{.12\textwidth}
      \includegraphics[width=\linewidth]{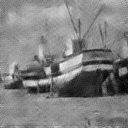}
     \end{minipage}
     &
    \begin{minipage}{.12\textwidth}
      \includegraphics[width=\linewidth]{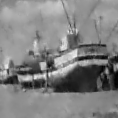}
     \end{minipage}
    \\   
    {\footnotesize Boat} & {\footnotesize $23.10\text{dB}/0.58$}  & {\footnotesize $24.78\text{dB}/0.68$} & {\footnotesize $23.50\text{dB}/0.68$}  & {\footnotesize $25.14\text{dB}/0.73$}  & {\footnotesize 		$\boldsymbol{26.91\text{dB}/0.89}$}  & {\footnotesize $26.61\text{dB}/0.86$} 
    \\ 
  \end{tabular}
  \label{BCNN_others}
\end{table*}

Fig. \ref{kld_bcnns}(a) shows that BCNN1 has the highest KLD (0.245). That means it has the worst ability to model the empirical distribution.
BCNN2 has smaller KLD (0.156) than BCNN1, since it enlarges $\mathcal{N}_1$ to $\mathcal{N}_2$ while remaining the value of $F_n$ and $\boldsymbol{\delta}$ the same as BCNN1. 
Increasing $F_n$ from 4 to 8 makes BCNN3 get smaller KLD (0.126) than BCNN2. 
Fig. \ref{kld_bcnns}(c) shows that the KLD (0.108) of BCNN4 is less than BCNN3 since the former has more GSM scales than the later.
BCNN5 achieves the smallest KLD (0.089), because we increase $\mathcal{N}_d$ and $F_n$ both.
In summary, the model with more hyper-parameters has smaller KLD, which means that it can simulate the empirical distribution better. 

Subsequently, the above BCNNs models are evaluated by grayscale image restoration, and SSM is selected as baseline here. 
Gaussian random matrix is used to obtain the measurement $\boldsymbol{y}$ from three standard test images (starfish, butterfly, and parrot), and image dimension is $128 \times 128$. 
Measurement ratio ($\text{MR} = |\boldsymbol{y}|/|\boldsymbol{x}|$) is 0.25, so the dimension of $\boldsymbol{y}$ is $4096 \times 1$, where $|\boldsymbol{x}| = 128^2$ is the cardinality of $\boldsymbol{x}$.
The performance of five BCNNs models are posted in Table \ref{BCNN_self}, 
which shows that all BCNNs models outperform SSM and the model with lower KLD achieves better restoration in most cases.

The superiority of BCNNs can be ascribed to three reasons.
First, BCNNs can derive optimal parameters by the efficient training algorithm.
For example, though BCNN1 employs the same $\mathcal{N}_1$ filters as SSM to simulate the cluster structure, BCNN1 can learn optimal convolutional filters $f_m(\cdot)$ via CD learning algorithm but SSM merely uses hand-crafted filters.
Moreover, BCNN1 can learn optimal weights $\pi_{mn}$ of GMM to preserve non-linearity precisely, which is also very difficult to realize by traditional Bayesian inference method \cite{xinjie}. 

Second, BCNNs can accommodate arbitrary convolutional filters in the architecture of CNNs.
More convolutional filters can describe more signal structures, thereby achieving better CS restoration.
But most prior distributions can only adopt few filters because of their limitations discussed in Section \ref{lim_bcs}.
For instance, SSM merely incorporates four linear filters to describe the cluster structure, which is not enough to achieve the state-of-the-art CS restoration.
In contrast, the number of linear filters in BCNN3 is twice as many as SSM.
Hence, BCNN3 can achieve better restoration.

\begin{figure}[!b]
\centering
\includegraphics[width = 0.99\linewidth]{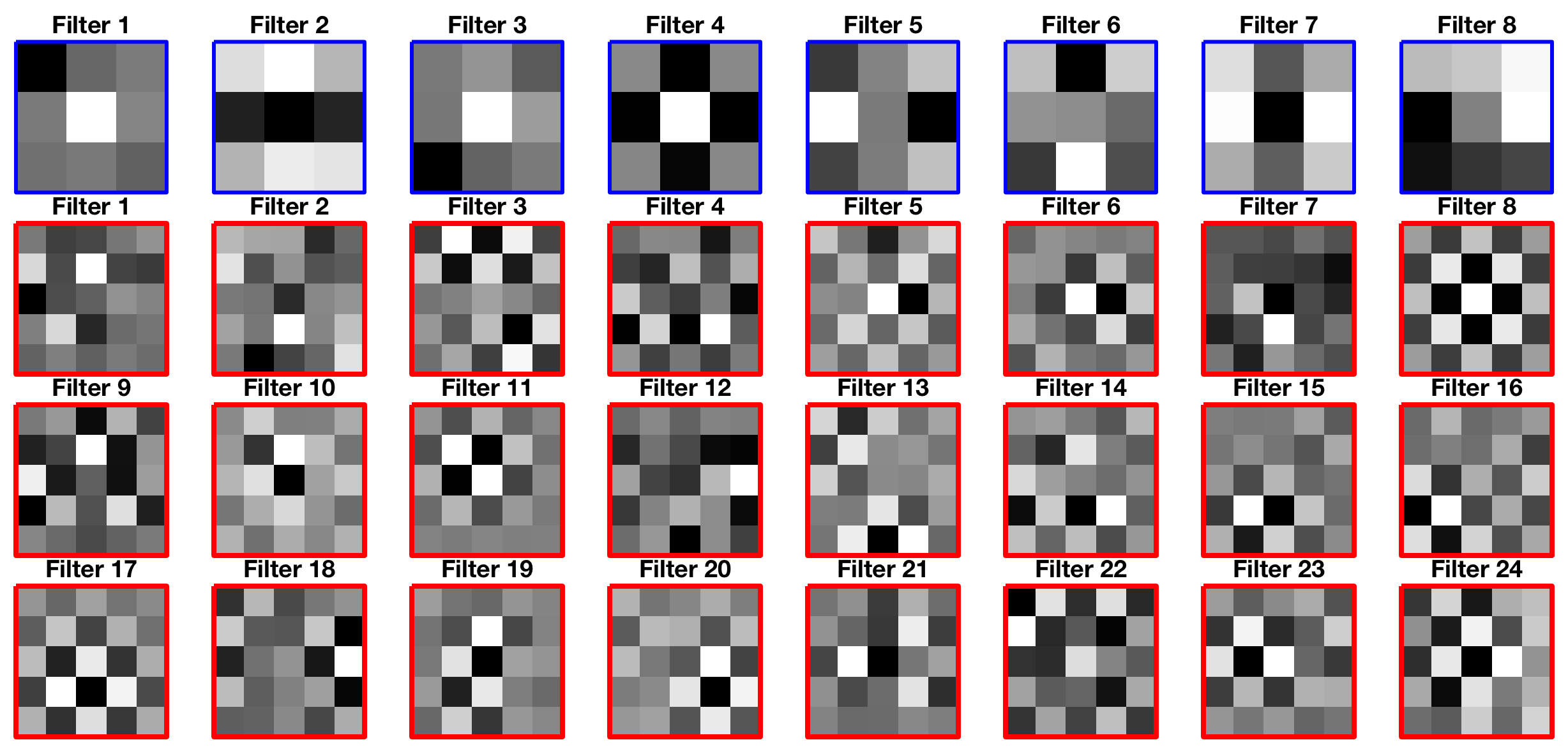}
\caption{The convolutional filters from BCNN4 and BCNN5. The first 8 filters with blue boundary belong to BCNN4, and the left 24 filters with red boundary belong to BCNN5.}
\label{conv_filters}
\end{figure}

Third, BCNNs can accommodate complicated convolution filters.
Most prior distributions can only model the signal structure in a small neighborhood, e.g.,
SSM merely describes the cluster structure in $\mathcal{N}_1$, which is hard to model high-order cluster structures. 
Compared to SSM, BCNNs can use more complicated filters to describe high-order structures.
Fig. \ref{conv_filters} draws all convolutional filters of BCNN4 and BCNN5, which shows that they can learn more complex signal structures than SSM.
That helps them to generate more powerful prior distributions and achieve much better CS restoration.

Notably, the superiority of BCNN5 over BCNN4 is not very striking, even though the former has more complex networks.
That is because BCNN4 has already captured most signal structures for CS restoration.
Though BCNN5 can learn more complex signal structures, but these learned signal structures can not provides much more information for CS restoration. 
Fig. \ref{conv_filters} shows that some filters of BCNN5 are similar as BCNN4, e.g., Filter 4 of BCNN4 and Filter 16 of BCNN5. 
Furthermoere, more complicated networks could make BCNN5 difficult to converge.

\begin{table}[!b]
         \centering
  	\caption{CS Restoration in Noiseless Situation}
  	\begin{tabular}{c c c c}
		\hline
		\hline
		\multirow{2}{*}{{\footnotesize Method}} & MR = 0.25 & MR = 0.10 & MR = 0.04 \\
		\cline{2-4}
	                 	& {\footnotesize PSNR(dB) SSIM} & {\footnotesize PSNR(dB) SSIM} & {\footnotesize PSNR(dB) SSIM} \\
		\hline
		{\footnotesize LASSO}	& 12.15\hspace{0.6cm}0.03	& 13.18\hspace{0.6cm}0.01	& 13.99\hspace{0.6cm}0.01 \\
        		{\footnotesize BCS}  		& 25.06\hspace{0.6cm}0.65	& 22.23\hspace{0.6cm}0.49 	& 20.12\hspace{0.6cm}0.37 \\
		{\footnotesize TV}       	& 26.20\hspace{0.6cm}0.71      & 20.73\hspace{0.6cm}0.44	& 19.39\hspace{0.6cm}0.26 \\
		{\footnotesize ReconNet} 	& 24.12\hspace{0.6cm}0.68      & 21.95\hspace{0.6cm}0.55	& 20.07\hspace{0.6cm}0.44 \\
		{\footnotesize DR2Net} 	& 25.86\hspace{0.6cm}0.74      & 22.77\hspace{0.6cm}0.59	& 20.42\hspace{0.6cm}0.46 \\
		{\footnotesize LDAMP} 	& 27.17\hspace{0.6cm}0.76	& 23.61\hspace{0.6cm}0.58	& 20.45\hspace{0.6cm}0.42 \\
		{\footnotesize BCNNs} 	& \textbf{28.58}\hspace{0.6cm}\textbf{0.81}      & \textbf{24.67}\hspace{0.6cm}\textbf{0.63}	& \textbf{21.27}\hspace{0.6cm}\textbf{0.47} \\
        		\hline
   	\end{tabular}
  \label{CS_25_Non_1}
\end{table}

\begin{table*}[h!]
  \centering
   \caption{Restoration Comparison [PSNR/SSIM] in Noisy Situation (SNR=12dB)}
  \begin{tabular}{ c c c c c c c}
    {\footnotesize Original} & {\footnotesize SSM} & {\footnotesize TV} & {\footnotesize ReconNet} & {\footnotesize DR2Net} & {\footnotesize LDAMP} & {\footnotesize BCNNs}
    \\ 
    \begin{minipage}{.12\textwidth}
      \includegraphics[width=\textwidth]{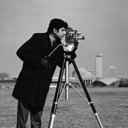}
    \end{minipage}
    &
      \begin{minipage}{.12\textwidth}
      \includegraphics[width=\textwidth]{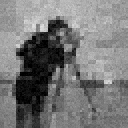}
     \end{minipage}
    & 
      \begin{minipage}{.12\textwidth}
      \includegraphics[width=\textwidth]{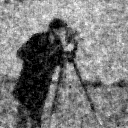}
     \end{minipage}
    &
    \begin{minipage}{.12\textwidth}
      \includegraphics[width=\textwidth]{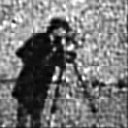}
     \end{minipage}
     &
    \begin{minipage}{.12\textwidth}
      \includegraphics[width=\textwidth]{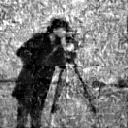}
     \end{minipage}
     &
    \begin{minipage}{.12\textwidth}
      \includegraphics[width=\linewidth]{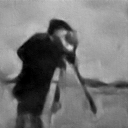}
     \end{minipage}
     &
    \begin{minipage}{.12\textwidth}
      \includegraphics[width=\linewidth]{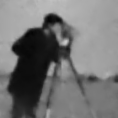}
     \end{minipage}
    \\   
    {\footnotesize Cameraman} & {\footnotesize $19.87\text{dB}/0.40$}  & {\footnotesize $19.36\text{dB}/0.33$} & {\footnotesize $18.39\text{dB}/0.39$}  & {\footnotesize $18.50\text{dB}/0.34$}  & {\footnotesize 	$21.42\text{dB}/0.63$}  & {\footnotesize $\boldsymbol{21.56\text{dB}/0.67}$} 
    \\ 
    \begin{minipage}{.12\textwidth}
      \includegraphics[width=\textwidth]{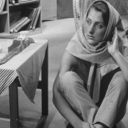}
    \end{minipage}
    &
      \begin{minipage}{.12\textwidth}
      \includegraphics[width=\textwidth]{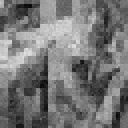}
     \end{minipage}
    & 
      \begin{minipage}{.12\textwidth}
      \includegraphics[width=\textwidth]{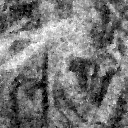}
     \end{minipage}
    &
    \begin{minipage}{.12\textwidth}
      \includegraphics[width=\textwidth]{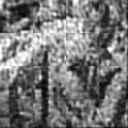}
     \end{minipage}
     &
    \begin{minipage}{.12\textwidth}
      \includegraphics[width=\textwidth]{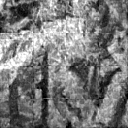}
     \end{minipage}
     &
    \begin{minipage}{.12\textwidth}
      \includegraphics[width=\linewidth]{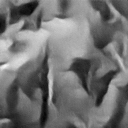}
     \end{minipage}
     &
    \begin{minipage}{.12\textwidth}
      \includegraphics[width=\linewidth]{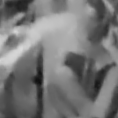}
     \end{minipage}
    \\   
    {\footnotesize Barbara} & {\footnotesize $19.71\text{dB}/0.38$}  & {\footnotesize $19.24\text{dB}/0.46$} & {\footnotesize $18.95\text{dB}/0.51$}  & {\footnotesize $19.07\text{dB}/0.49$}  & {\footnotesize 	$20.29\text{dB}/0.54$}  & {\footnotesize $\boldsymbol{21.30\text{dB}/0.58}$} 
    \\ 
    \begin{minipage}{.12\textwidth}
      \includegraphics[width=\textwidth]{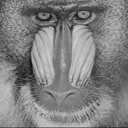}
    \end{minipage}
    &
      \begin{minipage}{.12\textwidth}
      \includegraphics[width=\textwidth]{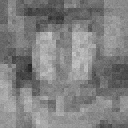}
     \end{minipage}
    & 
      \begin{minipage}{.12\textwidth}
      \includegraphics[width=\textwidth]{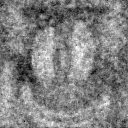}
     \end{minipage}
    &
    \begin{minipage}{.12\textwidth}
      \includegraphics[width=\textwidth]{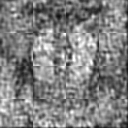}
     \end{minipage}
     &
    \begin{minipage}{.12\textwidth}
      \includegraphics[width=\textwidth]{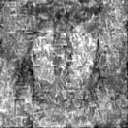}
     \end{minipage}
     &
    \begin{minipage}{.12\textwidth}
      \includegraphics[width=\linewidth]{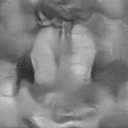}
     \end{minipage}
     &
    \begin{minipage}{.12\textwidth}
      \includegraphics[width=\linewidth]{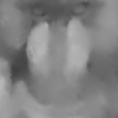}
     \end{minipage}
    \\   
    {\footnotesize Mandrill} & {\footnotesize $21.53\text{dB}/0.38$}  & {\footnotesize $19.78\text{dB}/0.36$} & {\footnotesize $18.53\text{dB}/0.32$}  & {\footnotesize $18.51\text{dB}/0.32$}  & {\footnotesize $22.08\text{dB}/0.43$}  & {\footnotesize $\boldsymbol{22.24\text{dB}/0.49}$} 
    \\ 
  \end{tabular}
  \label{BCNN_others_SNR12}
\end{table*}

\subsection{Comparisons with Other Algorithms in Noiseless Situation}

BCNN4 model is chosen to represent BCNNs algorithm for the following experiments, since it achieves the best tradeoff between network complexity and performance.
First, the above CS restoration methods are tested by three standard grayscale images (plane, peppers, and boat) with MR = 0.25 under the same experimental conditions as above.
Table \ref{BCNN_others} shows that BCNN4 outperforms the traditional CS algorithms, i.e., TV and SSM, due to the advantages of BCNNs discussed above.
BCNN4 also outperforms three DNNs algorithms, even though the former has much simpler architecture than the later.

To further validate BCNNs, all algorithms are evaluated by a testing dataset, which has 50 images randomly selected from Berkeley Segmentation Dataset and Benchmark (BSDS) dataset \cite{MartinFTM01}, and every image is cropped into $128 \times 128$.
Given three different MRs (MR = 0.25, 0.10, and 0.04), Table \ref{CS_25_Non_1} summarizes their average performances in noiseless situation.
It is noteworthy that BCNNs still outperform both traditional SCS and CNNs algorithms in all cases.

Also note that ReconNet does not outperform both SSM and TV algorithms, which validates the limitation of DNNs, i.e., simply applying DNNs to CS restoration cannot achieve the state-of-the-art performance, because CS measurement cannot provide enough prior knowledge for DNNs to formulate an expressive Bayesian hierarchical model. 

\subsection{Comparisons with Other Algorithms in Noisy Situation}

The CS restoration algorithms are first tested by three standard grayscale images (cameraman, barbara, and mandrill) in the noisy situation (SNR = 12dB) by adding Gaussian noise into CS measurement $\boldsymbol{y}$, and the other experimental conditions are the same as before.
Table \ref{BCNN_others_SNR12} shows their performances.
Futhermore, we use the same testing dataset from BSDS to further test these methods in two more noisy situations, i.e., SNR = 16dB and 8dB.
Their average performances are presented in Table \ref{CS_25_16_1} and Table \ref{CS_25_8}, respectively.

We can find that BCNNs still achieve the best performance in various noisy situations.
ReconNet and DR2Net do not outperform traditional BCS method in noisy situation, due to the vulnerability of DNNs. 
Since LDAMP and BCNNs are hybrid algorithms of DNNs and traditional BCS, they are more robust and achieve better performance in noisy situation.

\begin{table}[htp]
         \centering
  	\caption{CS Image Restoration with SNR=16dB}
  	\begin{tabular}{c c c c }
		\hline
		\hline
		\multirow{2}{*}{{\footnotesize Method}} & MR = 0.25 & MR = 0.10 & MR = 0.04 \\
		\cline{2-4}
	                 	& {\footnotesize PSNR(dB) SSIM} & {\footnotesize PSNR(dB) SSIM} & {\footnotesize PSNR(dB) SSIM} \\
		\hline
		{\footnotesize LASSO}	& 12.03\hspace{0.6cm}0.03	& 13.13\hspace{0.6cm}0.01	& 13.98\hspace{0.6cm}0.01 \\
        		{\footnotesize BCS}  		& 22.79\hspace{0.6cm}0.52	& 20.99\hspace{0.6cm}0.41 	& 19.34\hspace{0.6cm}0.32 \\
		{\footnotesize TV}       	& 22.86\hspace{0.6cm}0.51      & 19.89\hspace{0.6cm}0.38	& 17.02\hspace{0.6cm}0.24 \\
		{\footnotesize ReconNet} 	& 21.39\hspace{0.6cm}0.51      & 20.26\hspace{0.6cm}0.44	& 18.62\hspace{0.6cm}0.35 \\
		{\footnotesize DR2Net} 	& 21.93\hspace{0.6cm}0.52      & 20.77\hspace{0.6cm}0.46	& 18.91\hspace{0.6cm}0.37 \\
		{\footnotesize LDAMP} 	& 22.81\hspace{0.6cm}0.54	& 20.66\hspace{0.6cm}0.45	& 18.45\hspace{0.6cm}0.33 \\
		{\footnotesize BCNNs} 	& \textbf{23.54}\hspace{0.6cm}\textbf{0.64} & \textbf{21.96}\hspace{0.6cm}\textbf{0.51} & \textbf{19.95}\hspace{0.6cm}\textbf{0.40} \\
        		\hline
   	\end{tabular}
  \label{CS_25_16_1}
\end{table}

\begin{table}[htp]
         \centering
  	\caption{CS Image Restoration with SNR=8dB}
  	\begin{tabular}{c c c c }
		\hline
		\hline
		\multirow{2}{*}{{\footnotesize Method}} & MR = 0.25 & MR = 0.10 & MR = 0.04 \\
		\cline{2-4}
	                 	& {\footnotesize PSNR(dB) SSIM} & {\footnotesize PSNR(dB) SSIM} & {\footnotesize PSNR(dB) SSIM} \\
		\hline
		{\footnotesize LASSO}	& 11.43\hspace{0.6cm}0.02	& 12.85\hspace{0.6cm}0.01	& 13.83\hspace{0.6cm}0.01 \\
        		{\footnotesize BCS}  		& 20.61\hspace{0.6cm}0.38	& 19.20\hspace{0.6cm}0.29 	& 17.80\hspace{0.6cm}0.16 \\
		{\footnotesize TV}       	& 17.45\hspace{0.6cm}0.26      & 17.47\hspace{0.6cm}0.25	& 16.03\hspace{0.6cm}0.18 \\
		{\footnotesize ReconNet} 	& 16.08\hspace{0.6cm}0.29      & 16.35\hspace{0.6cm}0.28	& 15.18\hspace{0.6cm}0.23 \\
		{\footnotesize DR2Net} 	& 16.19\hspace{0.6cm}0.28      & 16.67\hspace{0.6cm}0.27	& 15.23\hspace{0.6cm}0.23 \\
		{\footnotesize LDAMP} 	& 18.89\hspace{0.6cm}0.33	& 17.07\hspace{0.6cm}0.30	& 16.58\hspace{0.6cm}0.22 \\
		{\footnotesize BCNNs} 	& \textbf{21.11}\hspace{0.6cm}\textbf{0.46} & \textbf{19.32}\hspace{0.6cm}\textbf{0.37} & \textbf{17.79}\hspace{0.6cm}\textbf{0.31} \\
        		\hline
   	\end{tabular}
  \label{CS_25_8}
\end{table}

\subsection{Comparisons with Deeper Neural Networks}

We design two new networks DR2Net-12 and LDAMP-12 to evaluate the effect of deeper refining networks.
Compared to the refining networks of DR2Net only consisting of six convolutional layers, the refining networks of DR2Net-12 includes twelve convolutional layers.
The difference between LDAMP-12 and LDAMP is the same.

Given the same experimental conditions as above, we presents the performances of DR2Net-12 and LDAMP-12 in Table \ref{CS_refining_networks} and Table \ref{CS_refining_networks_SNR12}.
We can find that DR2Net-12 shows similar performance as DR2Net, which further validates the limitation of DNNs for CS restoration.
Also note that LDAMP-12 achieves notable improvement over LDAMP and outperforms BCNNs in noiseless situation, but only shows similar performance as BCNNs in noisy situation.
In summary, BCNNs still demonstrate effective and robust CS restoration compared to deeper neural networks in both noiseless and noisy situations.

\begin{table}[htp]
         \centering
  	\caption{CS Restoration in Noiseless Situation}
  	\begin{tabular*} {0.49\textwidth}{@{\extracolsep{\fill}}  c c c c}
		\hline
		\hline
		\multirow{2}{*}{{\footnotesize Method}} & MR = 0.25 & MR = 0.10 & MR = 0.04 \\
		\cline{2-4}
	                 	& {\footnotesize PSNR(dB) SSIM} & {\footnotesize PSNR(dB) SSIM} & {\footnotesize PSNR(dB) SSIM} \\
		\hline
		{\footnotesize DR2Net}       		& 25.86\hspace{0.6cm}0.74      & 22.77\hspace{0.6cm}0.59	& 20.42\hspace{0.6cm}0.46 \\
		{\footnotesize DR2Net-\textbf{12}} 	& 26.49\hspace{0.6cm}0.76      & 22.97\hspace{0.6cm}0.61	& 20.47\hspace{0.6cm}0.48 \\
		{\footnotesize LDAMP} 			& 27.17\hspace{0.6cm}0.76	& 23.61\hspace{0.6cm}0.58	& 20.45\hspace{0.6cm}0.42 \\
		{\footnotesize LDAMP-\textbf{12}} 	& \textbf{29.58}\hspace{0.6cm}\textbf{0.82}	& \textbf{24.70}\hspace{0.6cm}0.63 	& \textbf{21.69}\hspace{0.6cm}\textbf{0.53}  \\
		{\footnotesize BCNNs} 			& 28.58\hspace{0.6cm}0.81      & 24.67\hspace{0.6cm}\textbf{0.68}	& 21.27\hspace{0.6cm}0.47 \\
        		\hline
   	\end{tabular*}
  \label{CS_refining_networks}
\end{table}

\begin{table}[htp]
         \centering
  	\caption{CS Restoration with SNR = 8dB}
  	\begin{tabular*} {0.49\textwidth}{@{\extracolsep{\fill}}  c c c c}
		\hline
		\hline
		\multirow{2}{*}{{\footnotesize Method}} & MR = 0.25 & MR = 0.10 & MR = 0.04 \\
		\cline{2-4}
	                 	& {\footnotesize PSNR(dB) SSIM} & {\footnotesize PSNR(dB) SSIM} & {\footnotesize PSNR(dB) SSIM} \\
		\hline
		{\footnotesize DR2Net}       		& 16.19\hspace{0.6cm}0.28      & 16.67\hspace{0.6cm}0.27	& 15.23\hspace{0.6cm}0.23 \\
		{\footnotesize DR2Net-\textbf{12}} 	& 16.37\hspace{0.6cm}0.28      & 16.52\hspace{0.6cm}0.27	& 15.19\hspace{0.6cm}0.22 \\
		{\footnotesize LDAMP} 			& 18.89\hspace{0.6cm}0.33	& 19.32\hspace{0.6cm}0.30	& 16.58\hspace{0.6cm}0.22 \\
		{\footnotesize LDAMP-\textbf{12}} 	& 20.81\hspace{0.6cm}\textbf{0.47}	& \textbf{19.47}\hspace{0.6cm}0.33	& 16.89\hspace{0.6cm}0.27 \\
		{\footnotesize BCNNs} 			& \textbf{21.11}\hspace{0.6cm}0.46 & 19.32\hspace{0.6cm}\textbf{0.37} & \textbf{17.79}\hspace{0.6cm}\textbf{0.31} \\
        		\hline
   	\end{tabular*}
  \label{CS_refining_networks_SNR12}
\end{table}

\section{Conclusion}

In this paper, we attempt to propose a statistical framework to explain the architecture of DNNs and design a novel algorithm for CS restoration based on the proposed framework.

The proposed framework of DNNs demonstrates that (i) neuron defines the energy of Gibbs distribution; (ii) the hidden layers of DNNs formulate Gibbs distributions; (iii) the whole architecture of DNNs can be explained as a BHM. 
In particular, this statistical framework provides a novel Bayesian perspective to understand the architecture of DNNs.

Based on the proposed framework, we provide insights into DNNs in four aspects: 
(i) we specify the application scope of DNNs; 
(ii) we reveal a limitation of DNNs for CS restoration; 
(iii) we find that an important reason for the vulnerability of DNNs is the activation function of DNNs cannot denoise very well; 
(iv) we unify traditional BCS methods and DNNs and provide an overall picture of all BCS algorithms.

We propose a novel CS restoration algorithm based on DNNs. 
In contrast to design an end-to-end DNNs for CS restoration, we propose a novel CNNs to model a prior distribution $p(\boldsymbol{x; \theta})$ only. 
The proposed CNNs have a dual property: (i) it has a closed-form expression as traditional prior distributions; 
(ii) it preserves the same powerful representation ability as DNNs. 
Based on the prior distribution $p(\boldsymbol{x; \theta})$ corresponding to the CNNs and the likelihood distribution of CS, we derive the posterior distribution of CS and propose a novel Bayesian inference algorithm for CS restoration.

Extensive simulations validate the superiority of BCNNs over the state-of-the-art CS restoration methods.
First, we show that BCNNs can learn a powerful prior distribution over traditional prior distributions, which is essential for traditional BCS algorithms.
Second, BCNNs show effective  performance than the existing CS restoration algorithms in noiseless and noisy situations. 
However, BCNNs are still not perfect. 
First, the speed of BCNNs is slower than the existing CS restoration algorithms, since the $p(\boldsymbol{x; \theta})$ corresponding to the CNNs are so complex that the posterior inference algorithm is time consuming.
Second, LDAMP with deeper refining networks can outperform BCNNs in noiseless situation. 

There are numerous directions for future work.
In terms of deep learning theory, the proposed framework of DNNs provides a new approach to solve the inherent problems of DNNs, such as proposing new DNNs architecture to defense the adversarial attacks in image classification field.
In the context of CS restoration, our future work can include the following topics.
First, we can refine the BCNNs algorithm to improve speed, which can extend the applications of BCNNs to real-time situations.
Second, we can propose new BCNNs architecture consisting of deeper neural networks to improve its performance further.


\section*{Acknowledgment}
The authors would like to thank Matiz Romero Sergio, Bayram Samet, and anonymous reviewers for their helpful comments and suggestions.
This work is supported by the National Science Foundation under Grant No. 1319598.

\ifCLASSOPTIONcaptionsoff
  \newpage
\fi


\bibliographystyle{IEEEtran}
\bibliography{IEEEabrv,cs-cnn}
%



%

\begin{IEEEbiography}[{\includegraphics[width=1in,height=1.25in]{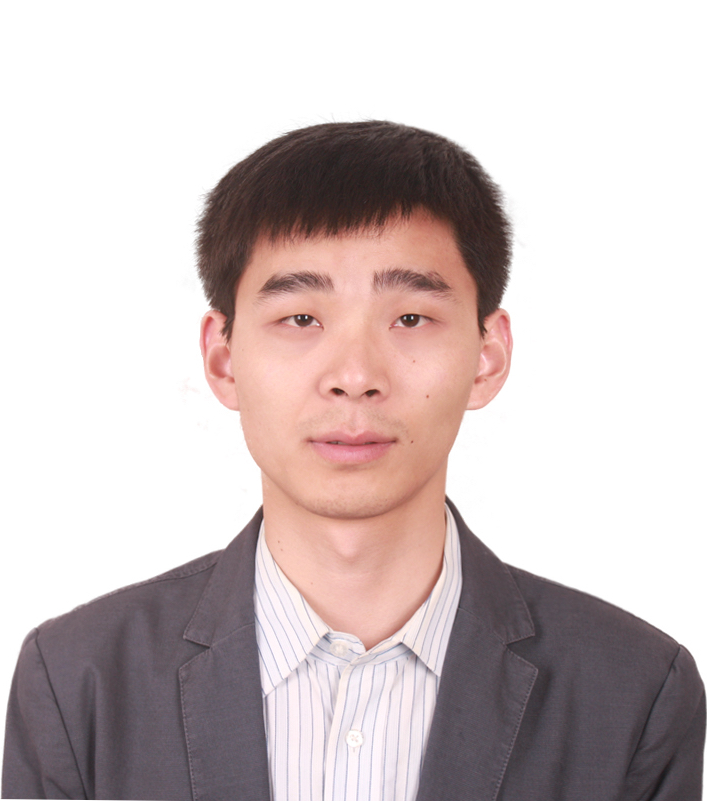}}]{Xinjie Lan}
(S'17) received the B.S.E.E. degree from Yantai University, Shandong, China, in 2010. He received the M.S.E.E. degree from Beijing Institute of Technology, Beijing, China, in 2012. From 2012 to 2014, He was an Assistant Engineer with Beijing Institute of Telemetry Technology, Beijing, China. He is currently a PhD candidate with the Department of Electrical and Computer Engineering, University of Delaware. His research interests include statistical signal processing, deep learning, inverse problem, compressed sensing, and image restoration.
\end{IEEEbiography}

\begin{IEEEbiography}[{\includegraphics[width=1in,height=1.25in]{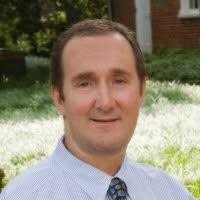}}]{Kenneth E. Barner}
(S'84-M'92-SM'00-F'16) \\ received the B.S.E.E. degree from Lehigh University, Bethlehem, Pennsylvania, in 1987. He received the M.S.E.E. and Ph.D. degrees from the University of Delaware, Newark, Delaware, in 1989 and 1992, respectively. He was the duPont Teaching Fellow and Visiting Lecturer with the University of Delaware in 1991 and 1992, respectively. From 1993 to 1997, he was an Assistant Research Professor with the Department of Electrical and Computer Engineering, University of Delaware, and a Research Engineer with the duPont Hospital for Children. He is currently Charles B. Evans Professor and Chairman with the Department of Electrical and Computer Engineering, University of Delaware. 

His research interests include signal and image processing, robust signal processing, nonlinear systems, sensor networks and consensus systems, compressive sensing, human-computer interaction. Dr. Barner is the recipient of a 1999 NSF CAREER award. He was the Co-Chair of the 2001 IEEE-EURASIP Nonlinear Signal and Image Processing (NSIP) Workshop and a Guest Editor for a Special Issue of the EURASIP Journal of Applied Signal Processing on Nonlinear Signal and Image Processing. He was a member of the Nonlinear Signal and Image Processing Board. He was the Technical Program Co-Chair for ICASSP 2005 and and previously served on the IEEE Signal Processing Theory and Methods (SPTM) Technical Committee and the IEEE Bio-Imaging and Signal Processing (BISP) Technical Committee. He is currently a member of the IEEE Delaware Bay Section Executive Committee. He has served as an Associate Editor of the IEEE Transactions on Signal Processing, the IEEE Transaction on Neural Systems and Rehabilitation Engineering, the IEEE Signal Processing Magazine, the IEEE Signal Processing Letters, and the EURASIP Journal of Applied Signal Processing. He was the Founding Editor-in-Chief of the journal Advances in Human-Computer Interaction. 
\end{IEEEbiography}




\end{document}